\documentclass{emulateapj}

\shorttitle{\emph{Spitzer} and FAST Survey of Cygnus X}
\shortauthors{Beerer et al.}
\usepackage{epsfig}
\slugcomment{Accepted for publication in The Astrophysical Journal}

\begin{document}

\title{A \emph{Spitzer} View of Star Formation in the Cygnus X North Complex} 

\author{ I.~M.~Beerer,\altaffilmark{1} X.~P.~Koenig,\altaffilmark{2}
  J.~L.~Hora\altaffilmark{2},
  R.~A. Gutermuth,\altaffilmark{3}\altaffilmark{4}
  S.~Bontemps,\altaffilmark{5} S.~T.~Megeath,\altaffilmark{6},
  N.~Schneider,\altaffilmark{5} F.~Motte\altaffilmark{7},
  S.~Carey,\altaffilmark{8} R.~Simon,\altaffilmark{9}
  E.~Keto,\altaffilmark{2} H.~A.~Smith,\altaffilmark{2}
  L.~E.~Allen,\altaffilmark{8} G.~G.~Fazio,\altaffilmark{2}
  K.~E.~Kraemer,\altaffilmark{11} S.~Price,\altaffilmark{11},
  D.~Mizuno,\altaffilmark{12} J.~D.~Adams,\altaffilmark{13}
  J.~Hern{\'a}ndez,\altaffilmark{14} P.~W.~Lucas\altaffilmark{15}}

\altaffiltext{1}{Department of Astronomy, University of California, Berkeley, CA, USA}
\altaffiltext{2}{Harvard-Smithsonian Center for Astrophysics, 60 Garden St., Cambridge, MA, USA}
\altaffiltext{3}{Smith College, Northampton, MA, USA}
\altaffiltext{4}{Dept. of Astronomy, University of Massachusetts, Amherst, MA, USA}
\altaffiltext{5}{Observatoire de Bordeaux, BP 89, 33270 Floirac, France}
\altaffiltext{6}{Dept. of Physics and Astronomy, University of Toledo, Toledo, OH, USA}
\altaffiltext{7}{AIM/SAp, CEA-Saclay, 91191 Gif Sur Yvette Cedex, France}
\altaffiltext{8}{Spitzer Science Center, Pasadena, CA, USA} 
\altaffiltext{9}{I. Physik. Institut, Universit\"{a}t zu K\"{o}ln, 50937 K\"{o}ln, Germany}
\altaffiltext{10}{NOAO, 950 North Cherry Avenue, Tucson, AZ, USA}
\altaffiltext{11}{Air Force Research Laboratory, Hanscom AFB, MA, USA}
\altaffiltext{12}{Institute for Scientific Research, Boston College, Boston, MA, USA}
\altaffiltext{13}{Cornell University, Department of Radiophysics Space Research, Ithaca, NY, USA} 
\altaffiltext{14}{Centro de Investigaciones de Astronom{\'i}a, Apdo. Postal 264, M{\'e}rida 5101-A, Venezuela}
\altaffiltext{15}{Centre for Astrophysics Research, Science \& Technology Research Institute, University of Hertfordshire, Hatfield, UK}

\begin{abstract}

  We present new images and photometry of the massive star forming
  complex Cygnus X obtained with the Infrared Array Camera (IRAC) and
  the Multiband Imaging Photometer for \emph{Spitzer} (MIPS) on board
  the \emph{Spitzer Space Telescope}. A combination of IRAC, MIPS,
  UKIRT Deep Infrared Sky Survey (UKIDSS), and Two Micron All Sky
  Survey (2MASS) data are used to identify and classify young stellar
  objects. Of the 8,231 sources detected exhibiting infrared excess in
  Cygnus X North, 670 are classified as Class I and 7,249 are
  classified as Class II. Using spectra from the FAST spectrograph at
  the Fred L. Whipple Observatory and Hectospec on the MMT, we
  spectrally typed 536 sources in the Cygnus X complex to identify the
  massive stars. We find that YSOs tend to be grouped in the
  neighborhoods of massive B stars (spectral types B0 to B9). We
  present a minimal spanning tree analysis of clusters in two regions
  in Cygnus X North. The fraction of infrared excess sources that
  belong to clusters with $\geq$10 members is found to be
  50--70$\%$. Most Class II objects lie in dense clusters within blown
  out \ion{H}{2} regions, while Class I sources tend to reside in more
  filamentary structures along the bright-rimmed clouds, indicating
  possible triggered star formation.

\end{abstract}


\section{Introduction}

High-mass stars, though relatively rare, dominate the energetics of
star formation in giant molecular clouds \citep{zinn07}. When
high-mass stars form through the gravitational collapse of gas and
dust, they release an immense amount of energy through stellar winds
and radiation. The release of this energy has a profound effect on the
massive star's environment, possibly inducing the formation of less
massive stars in the surrounding area. Therefore, regions that contain
massive stars provide excellent laboratories for studying both high
and low mass star formation. By studying the distribution of young
stellar objects (YSOs) in the neighborhood of massive stars, we hope
to understand the relationship between high-mass stars, star formation
and the interstellar medium (ISM).

Due to the high levels of extinction in star forming molecular clouds,
the use of infrared imaging instruments on the \emph{Spitzer Space
  Telescope} \citep{werner04} has provided great insight into regions
of star formation. \emph{Spitzer} has been used to survey a number of
star forming molecular clouds, e.g. W3 \citep{ruch07}, W5
\citep{koenig08} and M17 \citep{povich09,povich10}. In these surveys,
\emph{Spitzer} was used to detect infrared excess emission from warm
dust in disks and protostellar envelopes to map the spatial
distribution of low and high mass YSOs within the molecular clouds. In
this paper, we explore an active massive star-forming complex, Cygnus
X, using photometry from the \emph{Spitzer} Infrared Array Camera
(IRAC) \citep{fazio04} and the Multiband Imaging Photometer for
\emph{Spitzer} (MIPS) \citep{rieke04}.

Cygnus X is a $\sim 7\degr \times 7 \degr$ region in the Cygnus
constellation, which contains Gamma Cygni, the center star of the
Northern Cross \citep{reip08}. First known for being a diffuse radio
source, the region was named Cygnus X to distinguish it from the other
radio source in Cygnus, the radio galaxy Cygnus-A
\citep{piddington52}. Piddington and Minnett attributed the radio
source to thermal emission from clouds of ionized gas. Extensive radio
surveys of the region \citep[][and references therein and
subsequently]{wend84,wend91} have confirmed that thermal emission
dominates the radio wavelength output of Cygnus X, consistent with
Cygnus X being the largest and most active region of star formation
within 2~kpc of the Sun. It contains about 800 distinct \ion{H}{2}
regions, a number of Wolf-Rayet and OIII stars, several OB
associations, and at least 40 massive protostars. Cygnus X also
contains one of the most massive molecular cloud complexes in the
nearby Galaxy, with a mass of $\sim3\times 10^6 M_{\sun}$
\citep{schneider06}. Although once believed to be a superposition of
many disconnected star formation regions, Schneider et al. (2006,
2007) showed that the molecular clouds in Cygnus X form connected
groups at a distance of $\sim 1.7$ kpc. The entire region exhibits
evidence for many sites of star formation at different evolutionary
stages, from the youngest embedded star formation in Infrared Dark
Clouds (IRDCs) in DR21 \citep{downes66} to the young cluster Cygnus
OB2 \citep{knodl00} to the more dispersed and perhaps older Cygnus OB9
region.

None of the nearby Galactic regions (within 2~kpc of the Sun) studied
with \emph{Spitzer} are as richly populated with massive stars as
Cygnus X. Cygnus X is also near enough that the low-mass population
can be detected in the same \emph{Spitzer} observation. Observations
indicate that around 75\% of all stars form in clusters
\citep[][]{lada03,carpenter00,allen07}. Not all of the clusters in
these studies have high mass stars however. Other molecular cloud
surveys of massive star forming regions with \emph{Spitzer} have
revealed that 40--60\% of the low-mass stars do not form in
dense clusters \citep[e.g.][]{koenig08}. The identification and
classification of YSOs in Cygnus X enable us to study the spatial
distribution of these objects around massive stars. Determining how
many low mass stars form in clusters or in relative isolation will
help us understand how low-mass stars form in molecular cloud
complexes dominated by massive stars.

In this study, we used a combination of \emph{Spitzer} infrared
photometry and optical spectra from the Fast Spectrograph for the
Tillinghast Telescope (FAST) and Hectospec on the MMT to make a census
of the massive stars and the lower-mass YSOs in a $\sim 2\degr \times
2 \degr$ region, termed Cygnus X North, centered roughly on the
\ion{H}{2} region, DR21.

We describe our observations in $\S$~2. Our Cygnus X data were taken
with the \emph{Spitzer} IRAC and MIPS instruments as part of the
Cygnus X Legacy Survey program \citep{hora10}. The optical spectra
taken with the FAST and Hectospec instruments of stars in Cygnus X
North are presented in $\S$~2.4.

\begin{figure*}
\begin{center}
\epsfig{file=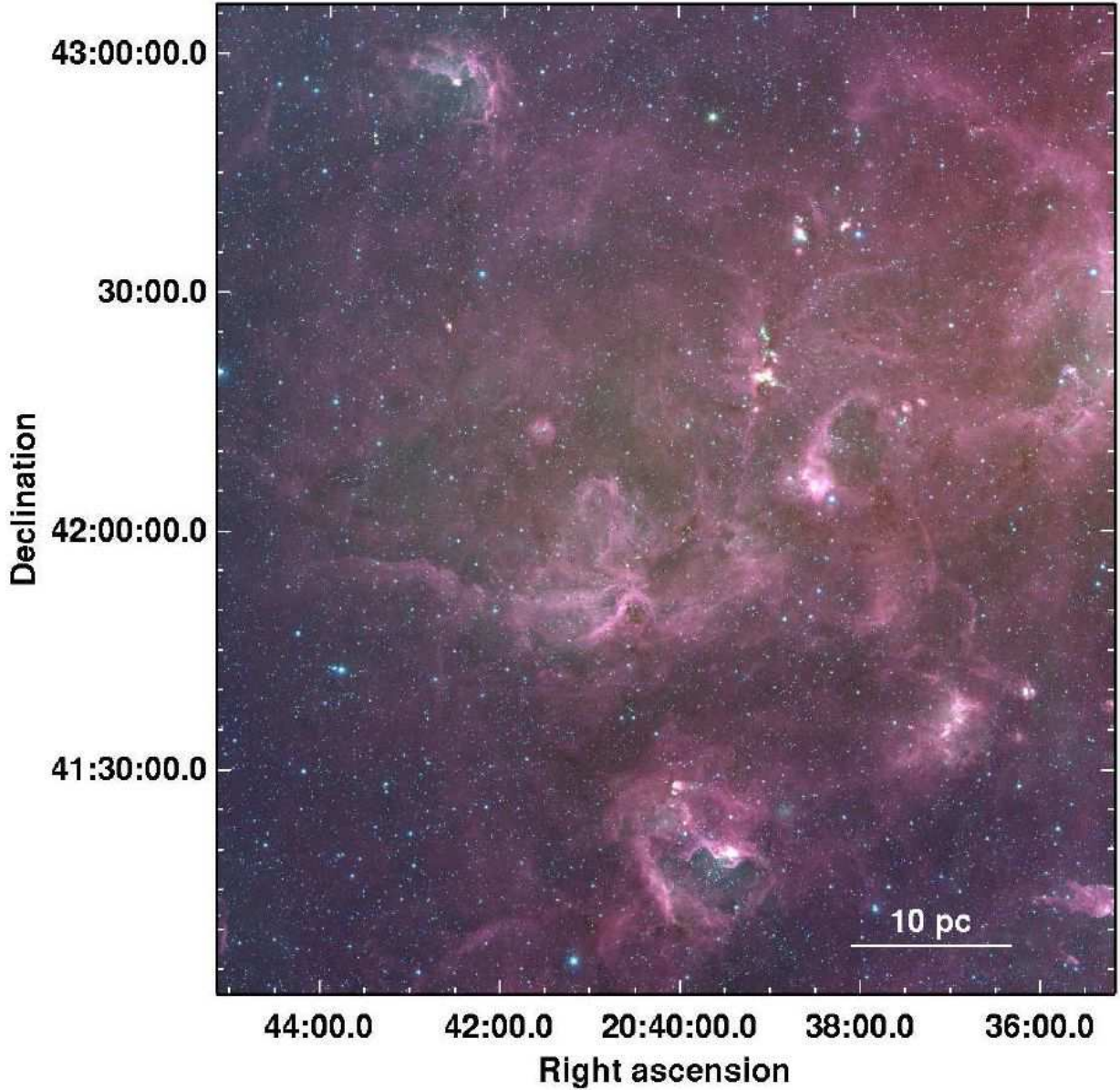,width=0.9\linewidth,clip=}
\caption{\emph{Spitzer} IRAC three-color composite image of Cygnus X
  North.~\label{fig-irac3} \emph{red}: 8.0 $\mu$m; \emph{green}: 4.5
  $\mu$m; \emph{blue}: 3.6 $\mu$m. The coordinate axes are in J2000.0
  epoch.}
\end{center}
\end{figure*}

\section{Observations and Methods}

\subsection{IRAC Data}

Cygnus X was observed with the \emph{Spitzer} IRAC instrument in all
four bands (3.6, 4.5, 5.8, and 8.0~$\micron$) in three campaigns
between 2007 November and 2008 November. A \emph{Spitzer} three-color
composite image of Cygnus X North is shown in
Figure~\ref{fig-irac3}. The RA and Declination for this and all
subsequent figures is J2000.0. The IRAC pixel scale is 1.2$\arcsec$
pixel$^{-1}$. The Cygnus X region was mapped by raster-scanning across
the field using offsets of 100$\arcsec$ in one direction and
300$\arcsec$ in the other, resulting in at least three frames at each
integration time at each position on the sky. The observations were
taken in HDR mode, in which a 10.4~s frame and a 0.4~s frame were
obtained at each position. Basic Calibrated Data (BCD) version 18.7
images from the \emph{Spitzer} Science Center's standard pipeline were
used. These data were combined into mosaics using WCSmosaic
\citep{gutermuth08}. The BCD images were improved by locating and
removing the column pull-down and mux-bleed artifacts and the banding
effects.

Automated source detection and aperture photometry were carried out on
all point sources using PhotVis version 1.10
\citep{gutermuth08}. PhotVis, a photometry visualization tool,
utilizes a DAOphot source-finding algorithm. The aperture photometry
was performed with synthetic apertures of 2.4$\arcsec$ radius and
background annuli of inner and outer radii of 2.4$\arcsec$ and
7.2$\arcsec$, respectively. The software filters outlier pixel values
in the background annuli via iterative statistics and adds the
measured standard deviation of the kept pixels in quadrature with the
shot noise estimates for the final mean background flux per pixel and
the shot noise in the main aperture. The photometry was calibrated
with the following values (Vega-standard magnitudes for 1 DN
s$^{-1}$): 19.455, 18.699, 16,498, and 16.892 for 3.6, 4.5, 5.8, and
8.0 $\mu$m bands, respectively. The completeness of the survey was
estimated by adding artificial sources of various magnitudes and
performing the photometry on the mosaics in the same way as was done
to generate the real catalog. The IRAC point spread function was used
to add the simulated sources into the images. In order for the source
to be recovered, it had to be within an arcsec of the position it was
inserted and within a half magnitude of the input value. Objects that
fall close to bright stars or bright, structured nebulosity were not
recovered. The 90$\%$ completeness levels for IRAC channels 1, 2, 3,
and 4 are 14.98, 14.87, 13.82, and 12.60, respectively
\citep{hora10}. Final photometric errors include the uncertainties in
the zero-point magnitudes ($\sim$0.03 mag).

\subsection{MIPS Data}

The MIPS observations were taken at the fast scan rate, with a typical
resolution of 6$\arcsec$ in the 24~$\mu$m band. Only the 24~$\mu$m
data are considered here. The data were reduced and mosaicked using
the MIPSGAL processing pipeline \citep{mizuno08, carey09}. Point
source photometry was extracted using Cluster Grinder and PhotVis
\citep{gutermuth08}. A synthetic aperture radius of 7.6$\arcsec$ and
background annulus inner and outer radii of 7.6$\arcsec$ and
17.8$\arcsec$ was used. The photometry was calibrated with the value
14.6 mag for 1 DN s$^{-1}$.

\begin{deluxetable}{lr}
\tablecaption{Source Classification Summary}
\tablewidth{0pt}
\tablehead{\colhead{Class} & \colhead{Number of Objects} } 
\startdata
Class I       &      670 \\
Class II      &      7,249 \\
Embedded protostars & 200 \\
Transition Disks    & 112 \\
Photospheres        & 350,058 \\
Other \tablenotemark{1}   &   677\\
Unclassified &  311,192 \\
Total        &  670,158\\
\enddata
\tablenotetext{1}{Includes PAH emission dominated sources,
  H$_2$ shock emission dominated sources and broad-line AGN
  candidates.\label{tbl-1} Unclassified sources lack detection in 4
  bands (either $HK_S$, IRAC 1 and 2, or IRAC 1, 2, 3 and 4) or a
  bright MIPS 24~$\micron$ detection.}
\end{deluxetable}

\subsection{YSO Classification}

Young stars emit excess radiation in the infrared compared with that
seen in main-sequence stellar photospheres, due to thermal emission
from the YSO's circumstellar material. Thus, YSOs can be identified by
looking for infrared excess emission.

YSOs are categorized into Class 0, I, II or III evolutionary
stages. Class 0 objects are deeply embedded protostars that are still
experiencing cloud collapse. Since they are so embedded, they are
extremely faint at wavelengths shorter than 10~$\micron$. They have a
significant submillimeter luminosity: $L_{\rm submm}/L_{\rm bol} >$
0.5\%. A Class I YSO is also an object whose emission is dominated by
a dense infalling spherical envelope but is infrared bright. With only
infrared data it is usually not possible to distinguish Class 0 from
Class I objects. A Class II YSO is characterized by the presence of an
optically thick, primordial circumstellar disk, which dominates the
star's emission. When most of the circumstellar disk material has
become optically thin, the star is classified as a Class III
star. Thus, a Class III object is a pre-main sequence star that has
lost its accretion disk, but may exhibit a small amount of infrared
excess due to a secondary/debris disk. In this paper we additionally
catalog a further class of YSO, the ``transition disk'' (TD). These
objects display an intermediate infrared excess morphology between
Class II and III as they exhibit disks, but with evidence for
significant dust clearing in the inner disk region. They exhibit
emission consistent with reddened photospheres that have IR excess at
24~$\micron$ and longer wavelengths. There is some debate in the
literature as to their exact place in the evolutionary path from Class
II to Class III however \citep[e.g.][]{hern08,currie10}.

The circumstellar material around a young star disappears as the star
evolves, and consequently the infrared excess emission
decreases. Thus, by measuring the star's IR excess---once it has
become visible in the infrared---we can determine its evolutionary
state. Since younger stars are brighter at longer IR wavelengths, we
can classify YSOs by calculating the slope of the source's spectral
energy distribution (SED), which is defined as:

\begin{equation}
\alpha=\frac{\emph{d} \log(\lambda F_\lambda)}{\emph{d} \log(\lambda)}
\end{equation}
where $F_\lambda$ is the star's flux density at wavelength, $\lambda$.

\begin{figure} 
\epsscale{2.0}
\plottwo{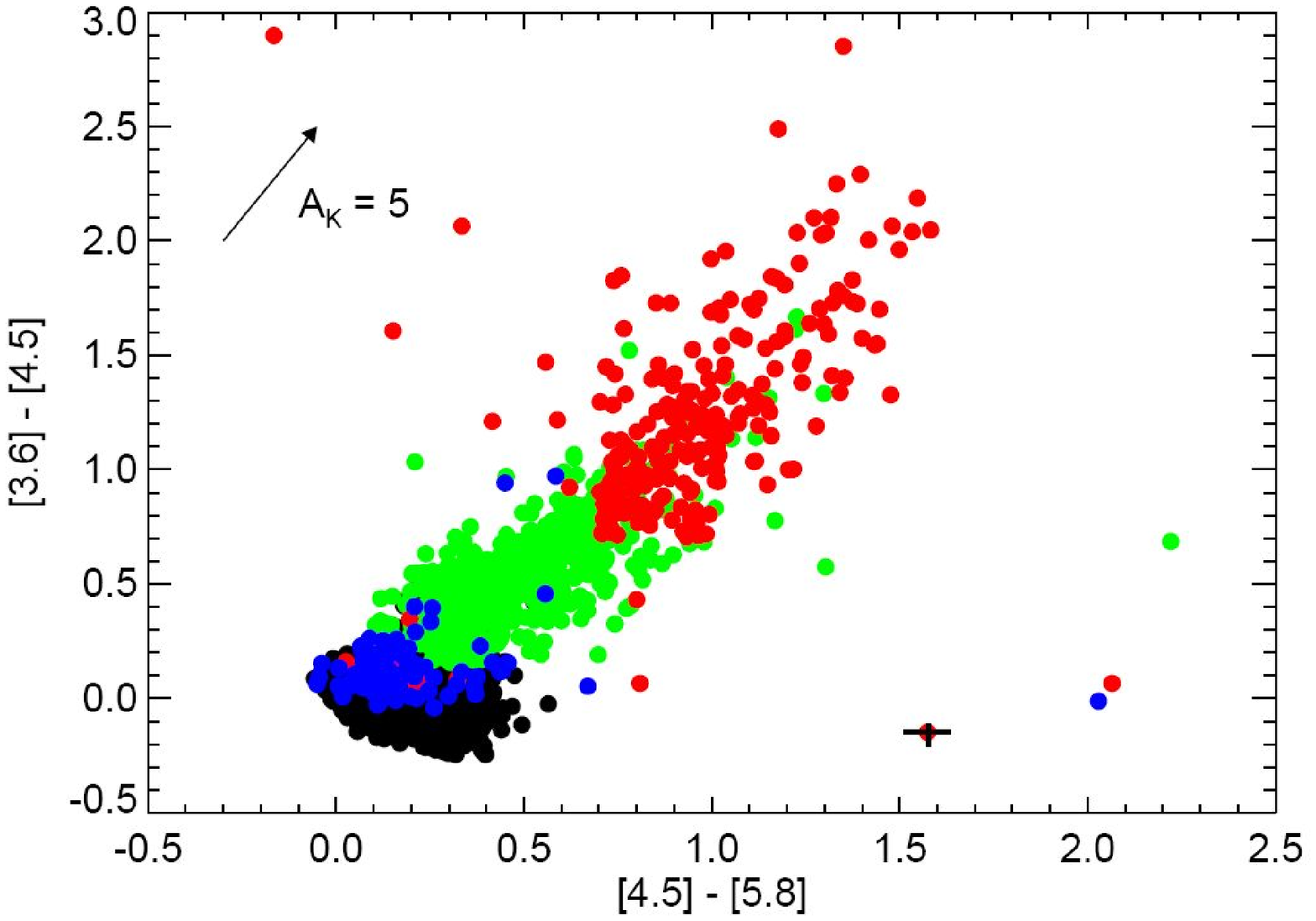}{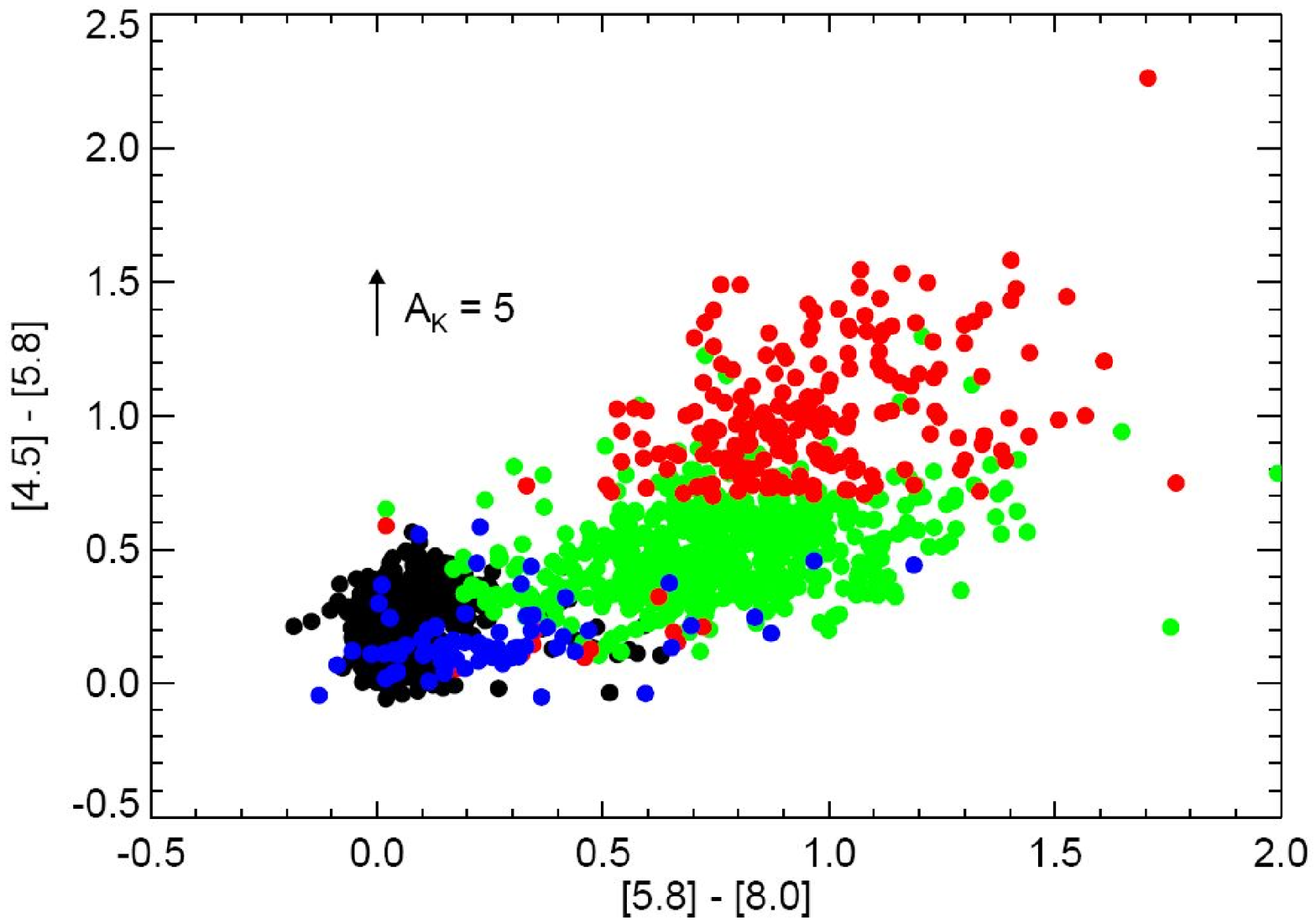}
\caption{\emph{Top}: [3.6]-[4.5] vs. [4.5]-[5.8]; \emph{Bottom}:
  [4.5]-[5.8] vs. [5.8]-[8.0] IRAC color-color diagrams used for
  identifying YSOs.\label{fig-ccdirac} One red dot shows error bars in
  black. \emph{Black dots}: Photospheres; \emph{green}: Class II;
  \emph{red}: Class I; \emph{blue}: Transition Disk Candidates.}
\end{figure}

\begin{figure}
\epsscale{2.0}
\plottwo{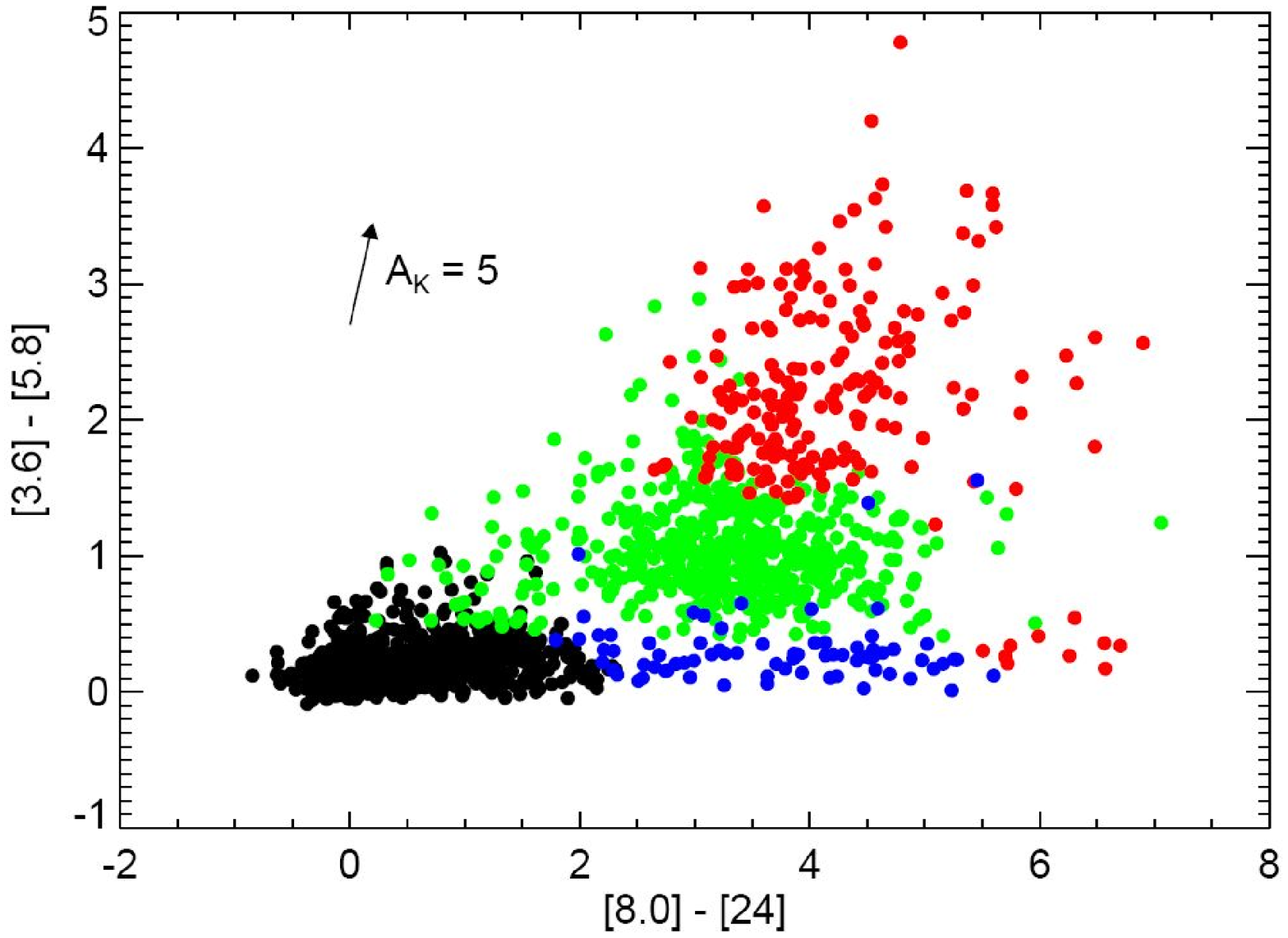}{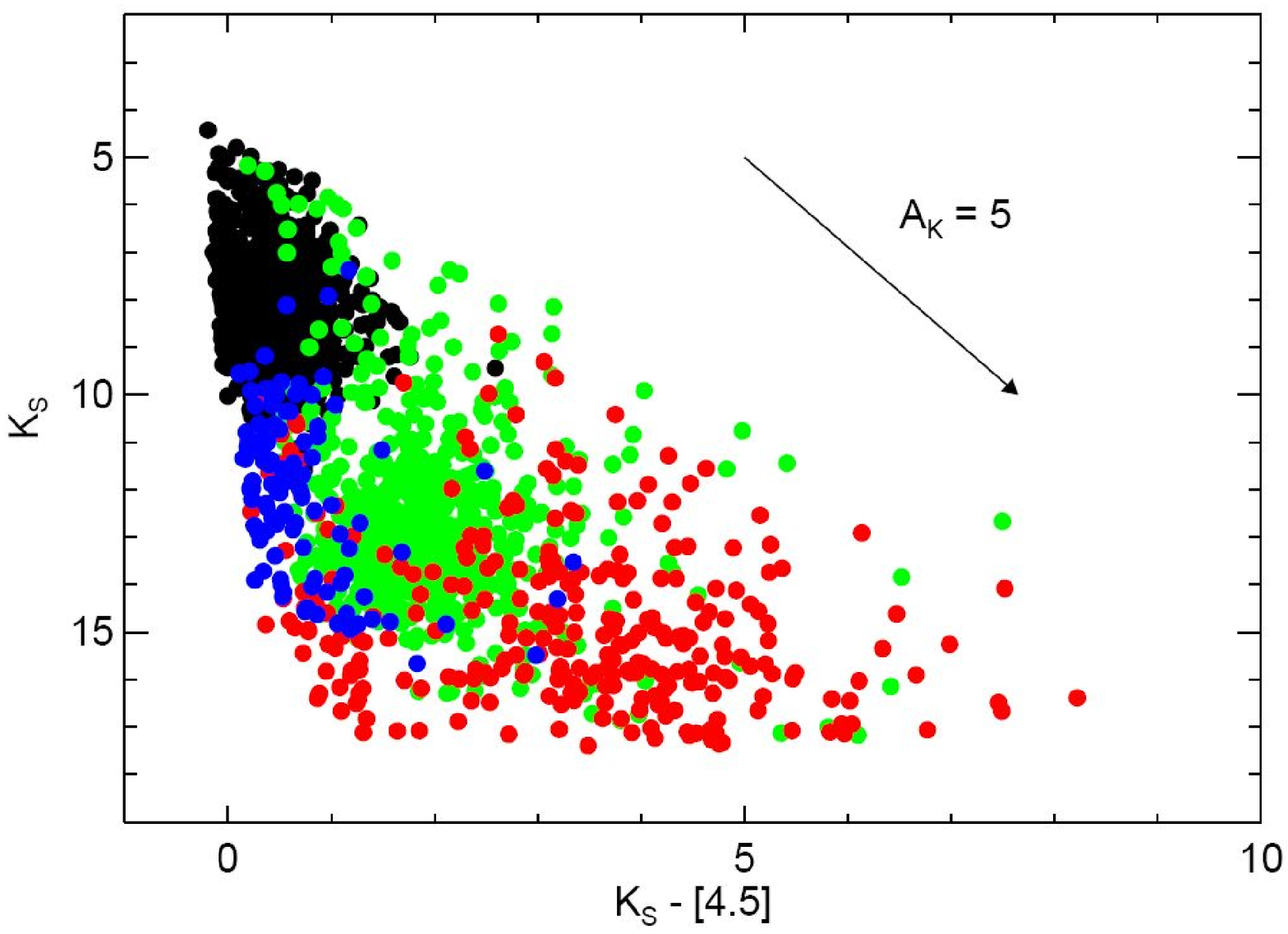}
\caption{\emph{Top}: [3.6]-[5.8] vs. [8.0]-[24] IRAC and MIPS
  color-color diagram; \emph{Bottom}: 2MASS [K] vs. [K]-[4.5]
  color-magnitude diagram used for identifying YSOs.\label{fig-ccdmip}
  \emph{Black dots}: Photospheres; \emph{green}: Class II; \emph{red}:
  Class I; \emph{blue}: Transition Disk Candidates.}
\end{figure}

Color-color and color-magnitude diagrams, which compare the ratio of a
star's flux in different bands, can be used to look for IR excess to
classify YSOs in place of calculating $\alpha$ for each object. We
used IRAC, MIPS, Two-Micron All-Sky Survey \citep[2MASS,][]{skrut06}
and UKIRT Deep Sky Survey DR4 \citep[UKIDSS,][]{law07, lucas08}
photometry to classify the YSOs in Cygnus X using the classification
scheme presented in \citet{gutermuth08}, with the extinction map of
\citet{schneider06} generated from 2MASS data used to deredden the
photometry. Figures~\ref{fig-ccdirac} and \ref{fig-ccdmip} show
example color-color and color-magnitude diagrams that were used to
classify the YSOs. The points are plotted without dereddening their
photometry. The scheme also provides a means of filtering out
extragalactic sources (for example, active galactic nuclei or AGN) on
the basis of magnitude and color cuts.

\begin{figure*}
\centering
\begin{tabular}{cc}
\epsfig{file=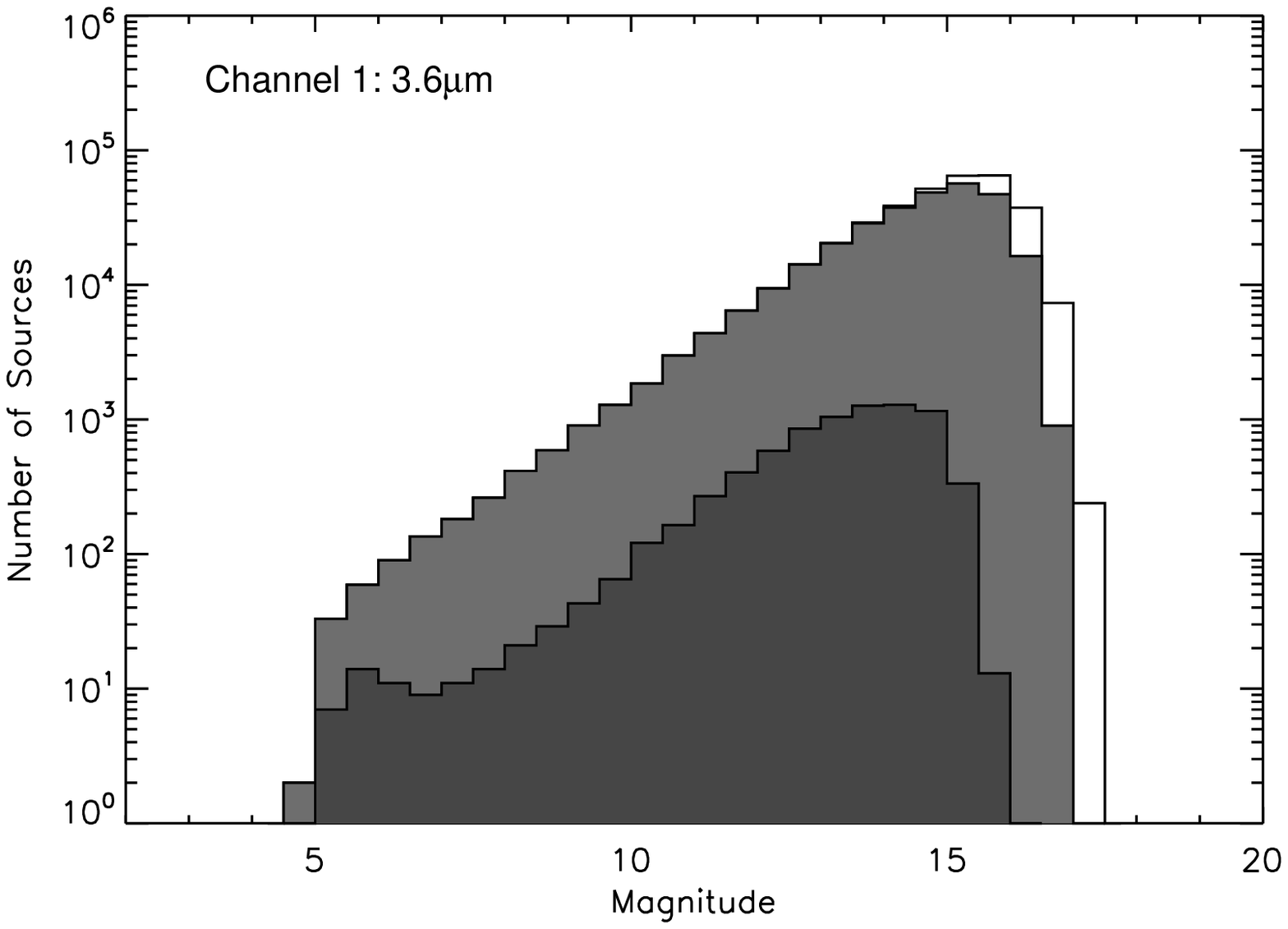,width=0.45\linewidth,clip=}&\epsfig{file=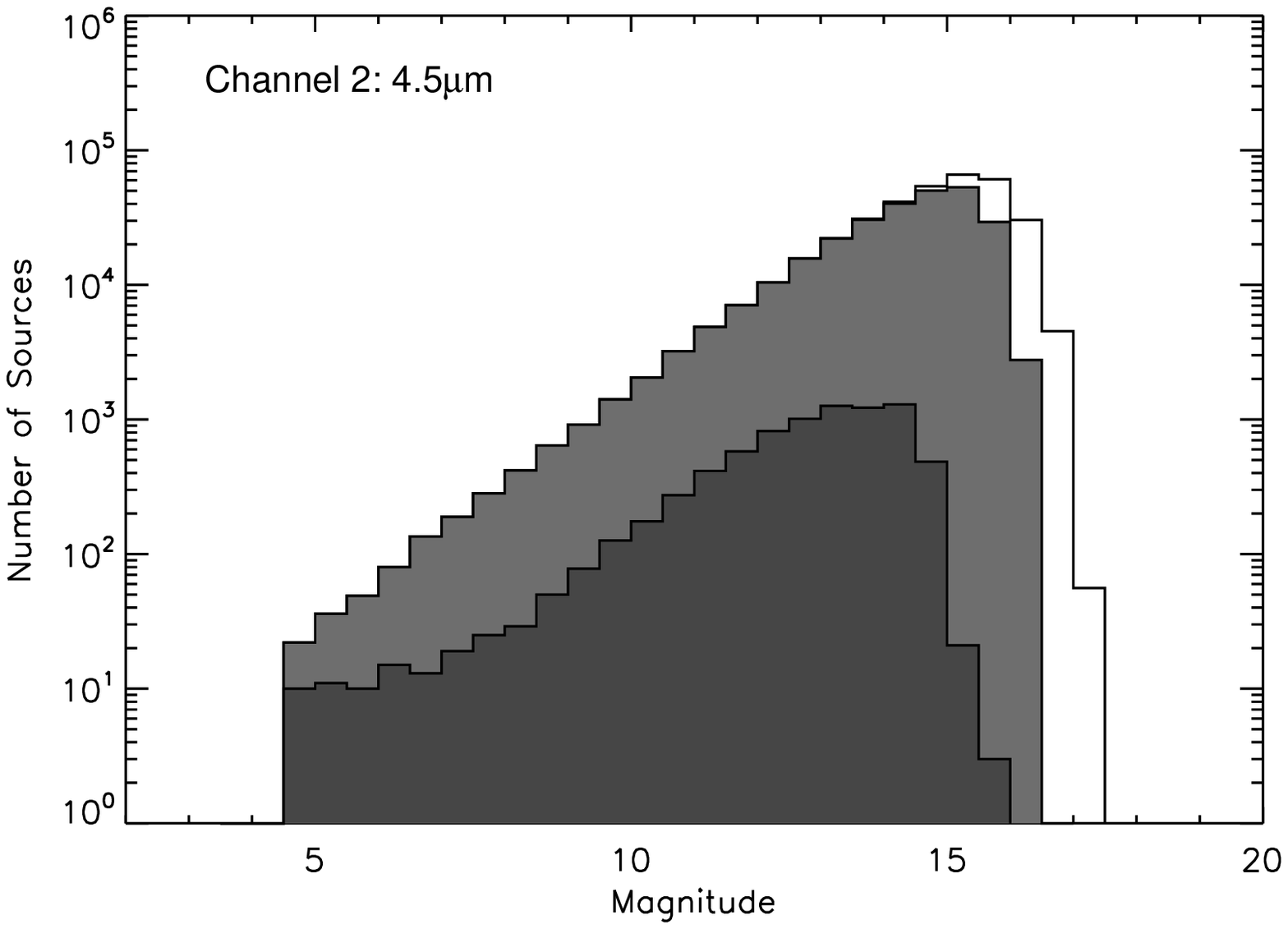,width=0.45\linewidth,clip=}\\
\epsfig{file=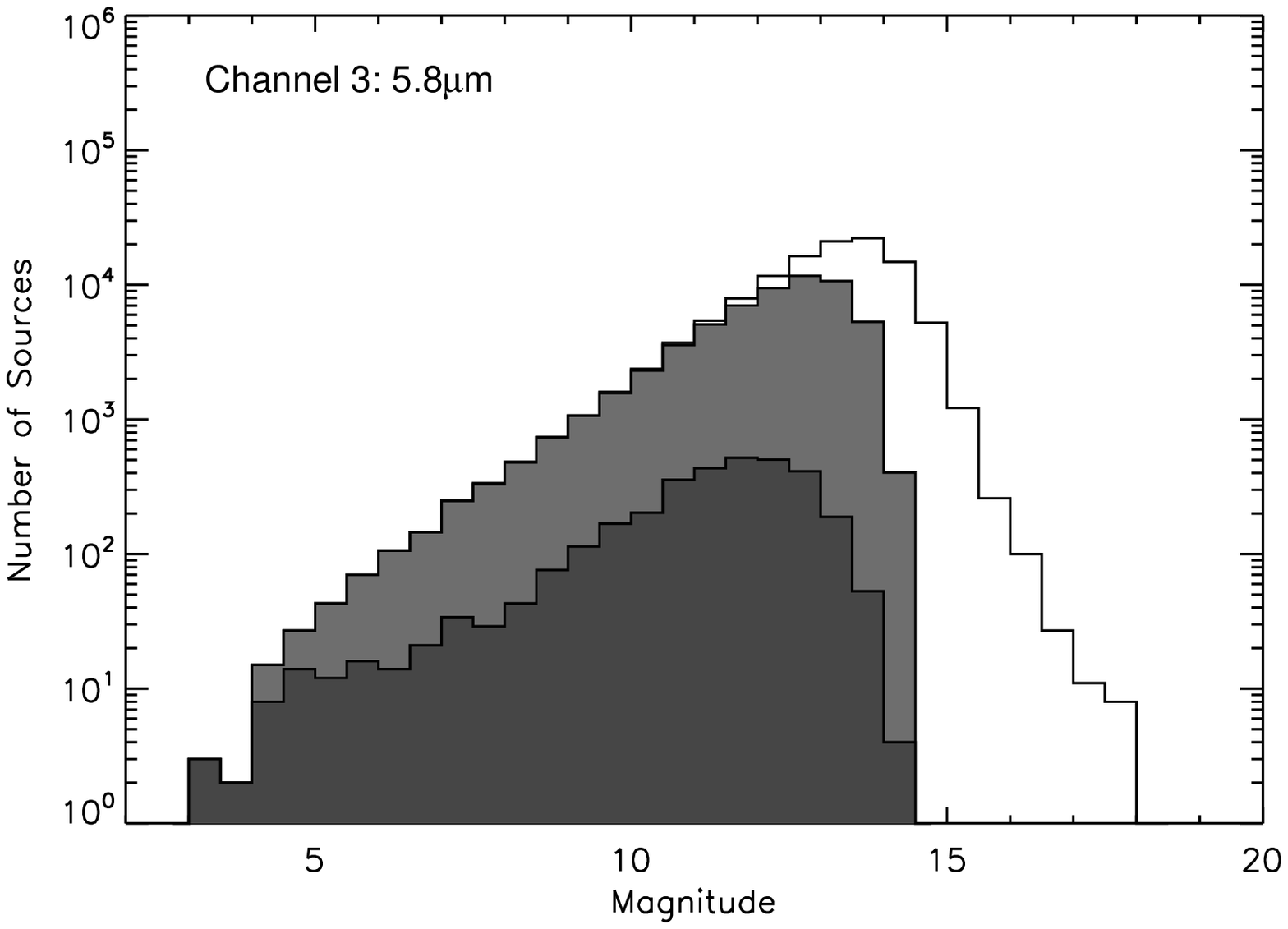,width=0.45\linewidth,clip=}&\epsfig{file=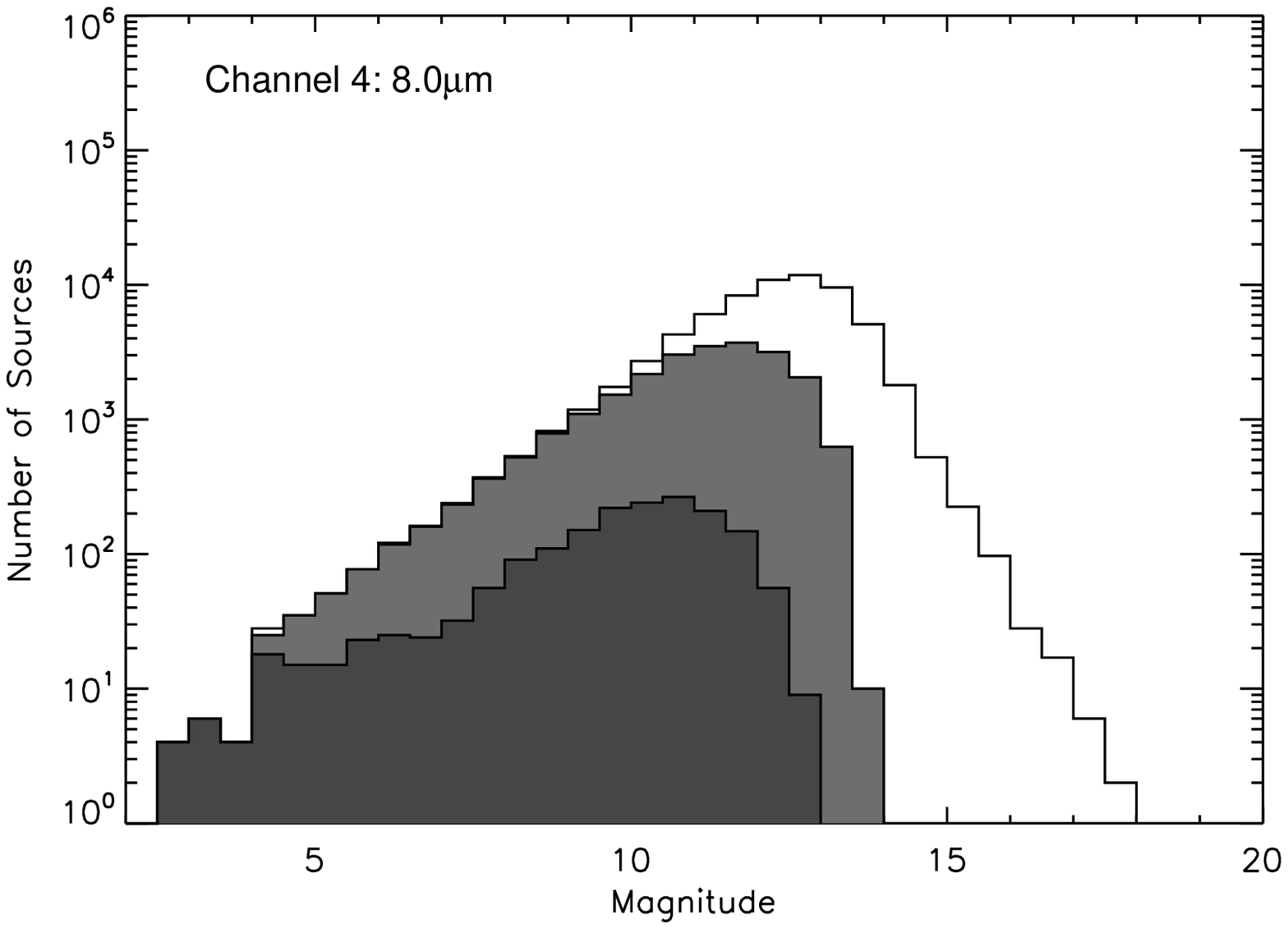,width=0.45\linewidth,clip=}
\end{tabular}
\caption{Apparent magnitude histograms of detections for all IRAC
  bands of all stellar sources (white histogram), stellar sources with
  error $<$0.1 mag (light gray histogram) and YSOs with error $<$0.1
  mag (dark gray histogram) in Cygnus X North.\label{fig-histoyso}}
\end{figure*}

Using these methods, we found 670 Class I, 7,249 Class II and 350,058
normal photospheres in the Cygnus X North region. We also identified
200 deeply embedded objects (with SEDs resembling Class I sources but
showing evidence for large optical extinction) and 112 transition
disks. These classifications are summarized in Table~\ref{tbl-1}.

In the left panel of Figure~\ref{fig-ccdirac}, one red Class I source
shows error bars. We calculated the errors in color by adding the
median uncertainty in magnitude for the two channels in quadrature. We
also show the extinction vector in each plot for that filter
combination using the extinction law derived by
\citet{flaherty07}. The typical line of sight extinction to sources in
Cygnus X is $A_K$ = 0.5 to 1.0 \citep{schneider06}. Thus, except for
the most embedded sources, neither photometry error or reddening are
responsible for the distribution of sources on these color-color
diagrams and the dominant contributor to their colors is the intrinsic
spread in properties of the sources. In Figure~\ref{fig-histoyso}, we
show the number of YSOs (including Class I, II, deeply embedded
protostars and transition disks) detected in each of the four IRAC
bands per magnitude, in comparison to all detected stellar
sources. Firstly we note that channels 3 and 4 detect far fewer
sources than channels 1 and 2. This deficit exists because channels 1
and 2 are more sensitive than channels 3 and 4 and are also less
affected by the bright diffuse emission that dominates the channel 3
and 4 images. The underlying typical stellar photosphere is also
intrinsically fainter at 5.8 and 8~$\micron$ than at 3.6 and
4.5~$\micron$, which further hinders our ability to detect sources in
the longer wavelength IRAC filters. Secondly, there is a flattening
off in the YSO histograms at the brightest magnitudes in all bands. We
attribute this trend to contamination, likely from AGB stars. As found
by \citet{robitaille08}, a majority of objects brighter than [4.5] =
7.8 in a typical Galactic field are so-called $``$extreme$''$ AGB
stars, while sources fainter than this cut are either $``$Standard$''$
AGB stars or YSOs. In Cygnus X this contamination reveals itself in
the apparent increased proportion of YSOs amongst the overall stellar
populations in each IRAC histogram at the brightest magnitudes
(brighter than 8th magnitude in any band).

\subsection{Optical Spectroscopy}
 
\begin{figure}
\begin{center}
\includegraphics[scale=1.15]{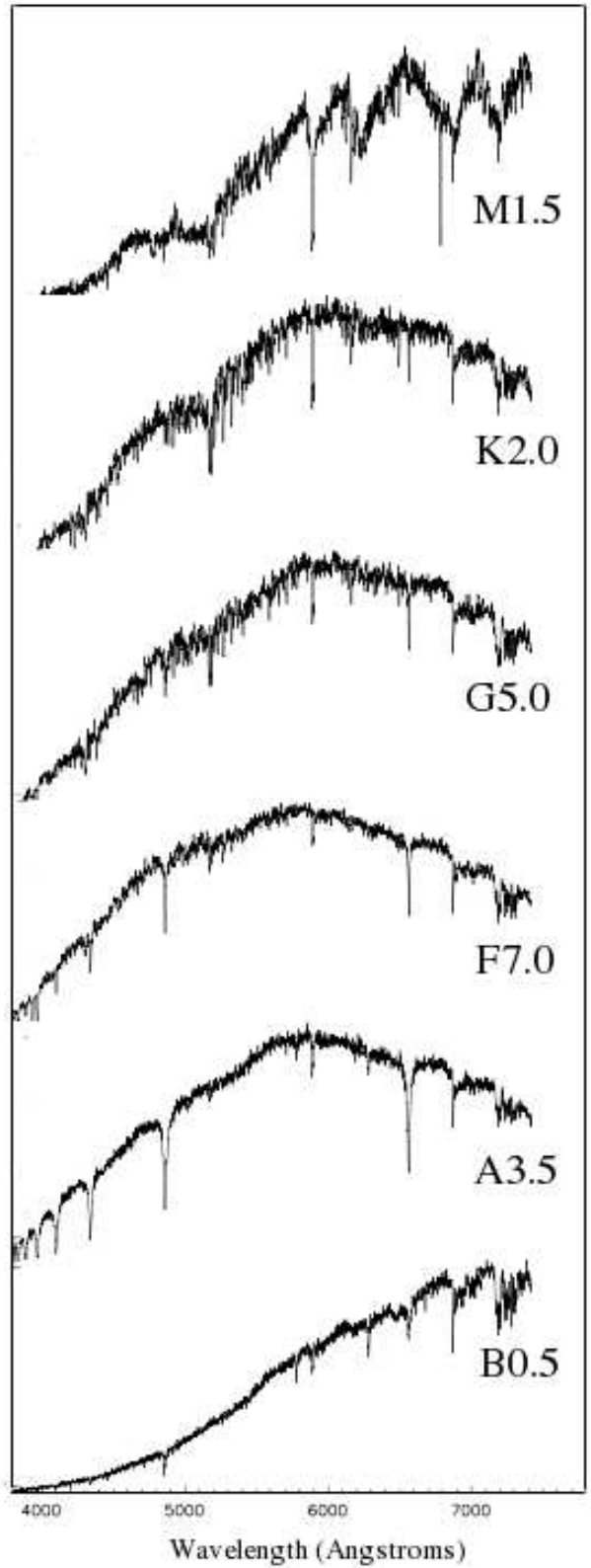}
\caption{Examples of our FAST optical spectra for each spectral
  type.\label{fig-specex}}
\end{center}
\end{figure}

To study the effect massive stars have on their environment in Cygnus
X, we need to first identify the most massive stars. Then, we can
investigate how the young stellar objects form in their
presence. Using $r^\prime$ and $i^\prime$ magnitudes from the IPHAS
survey \citep{drew05}, we identified a sample of possible O and B type
stars in the 2$\degr \times 2 \degr$ field near DR21 in Cygnus X
\citep{gonzalez08}. We obtained 536 optical spectra during Fall 2008
using the FAST instrument, an optical spectrograph in operation at the
focus of the 1.5~m Tillinghast reflector at the Fred L. Whipple
Observatory on Mt. Hopkins \citep{fab98} and the Hectospec multifiber
spectrograph mounted on the 6.5~m MMT telescope on Mount Hopkins
\citep{fab94}. The FAST spectra were obtained using the 300 groove
mm$^{-1}$ grating centered at 5500 {\AA} with a resolution of 3 {\AA}
and processed using the standard FAST pipeline reduction
\citep{tokarz97}. Figure~\ref{fig-specex} presents example FAST
spectra from our sample.

Hectospec is a multi-object spectrograph with 300 fibers that can be
placed within a 1{\degr} diameter circular field. We used the
270~groove mm$^{-1}$ grating, and obtained spectra in the range
3700--9000~{\AA} with a resolution of 6.2~{\AA}. Data reduction was
performed by S. Tokarz through the CfA Telescope Data Center, using
IRAF tasks and other customized reduction scripts. The reduction
procedure was the standard for the Hectospec data, with the addition
of our special sky subtraction procedure. To avoid the difficulties
encountered when subtracting a sky spectrum made from an average over
the highly variable H\,{\sc ii} region, we took exposures offset by
$\sim$5{\arcsec} after each Hectospec configuration. This way we
obtained a sky spectrum very close to each star through the same
fiber. Background subtraction was performed in IDL to remove each
wavelength calibrated sky offset spectrum from its corresponding
wavelength calibrated object spectrum.

\begin{figure}
\begin{center}
\includegraphics[scale=0.5]{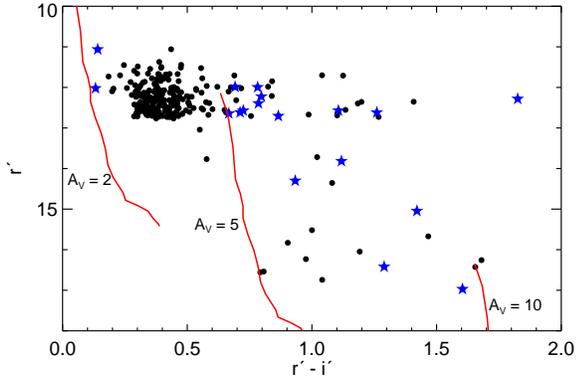}
\caption{Color-magnitude diagram of the spectroscopic sample in this
  paper.\label{fig-cmd1} Blue star symbols mark objects classified as
  B stars. Solid red lines show the main sequence between B0 and B9 at
  a distance of 1.7~kpc at $A_V$=2, 5 and 10 from Kenyon \& Hartmann
  (1995). Black dashed line shows expected locus of B0V stars at a
  distance of 1.7~kpc over the same range of optical extinction.}
\end{center}
\end{figure}

To classify our sources, we used the spectral classification code
SPTclass\footnote{http://www.astro.lsa.umich.edu/$\sim$hernandj/SPTclass/sptclass.html.}
\citep{her04, her05}. This automated code can classify spectra to a
precision of half a subtype (e.g. G6.5$\pm$0.5). It uses only spectral
line ratios, not continuum shape and is currently unable to
distinguish the luminosity class of the star being classified. Of the
536 spectra, 24 sources were classified as B type stars. The most
massive star identified was a B0.0$\pm$2.0. The fact that we did not
find any O type stars may be a result of our selection criteria. We
selected sources from the IPHAS survey, which does not include any
sources brighter than magnitude 10. Therefore, our survey may be
missing some of the brightest sources. Figure~\ref{fig-cmd1} shows the
optical color-magnitude distribution of IPHAS photometry for sources
for which we obtained spectra. The region above the black dashed line
is the expected location of O stars at the distance of Cygnus X
(1.7~kpc). Although we have some spectra in this region, the sampling
is sparse. We searched the literature for massive stars that were
identified in previous studies. \citet{comeron08} found one object
(star 100 in their catalog) classified as O9V in the vicinity of the
Diamond Ring (DR~17). They also identify many candidate early-type
stars in this region, although these lack a precise spectral
type. \citet{wend91} discuss the radio continuum emission at 408 and
4800~MHz in the environment of this portion of Cygnus X and show that
it is made up of numerous thermally emitting \ion{H}{2} regions, that
are likely powered by massive O stars. We thus attribute our lack of
confirmed spectroscopic O stars to a simple sampling
effect. Table~\ref{tbl-2} summarizes our spectral classification
results. The drop in number of objects classified later than spectral
type F (G, K and M) is due to the spectral sample selection which
preferentially chose the bright, blue targets and the sensitivity
limits of our spectral survey. The full listing of sources for which
we determined classifications, along with their positions, spectral
types and uncertainties is presented in the Appendix.

\begin{deluxetable}{cc}
\tablecaption{Spectral Classification Summary}
\tablewidth{0pt}
\tablehead{\colhead{Spectral Type} & \colhead{Number} }
\startdata
B &   24 \\
A &   49 \\
F &  229 \\
G &  146 \\
K &   81 \\
M &    7 \\
\label{tbl-2}
\enddata
\end{deluxetable}

\section{Analysis}

\subsection{Color-Magnitude Diagram}

\begin{figure}
\begin{center}
\includegraphics[scale=0.80]{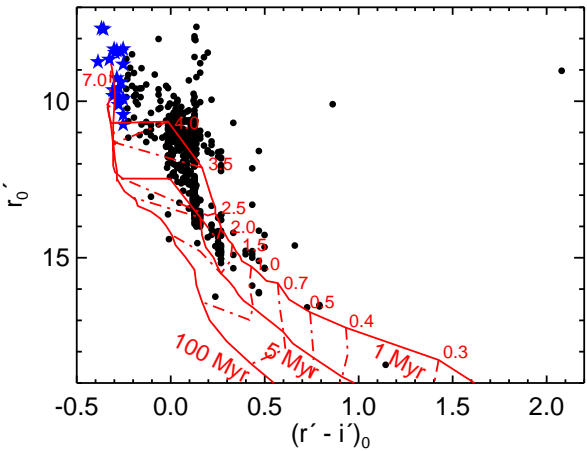}
\caption{Dereddened optical color-magnitude diagram of all sources for
  which we have optical spectral classifications.\label{fig-cmd2} B
  type stars are plotted with blue star symbols. A through M type
  stars are plotted in black. Red solid lines show isochrones for 1
  Myr, 5 Myr, and 100 Myr from Siess et~al. (2000). Red dash-dotted
  lines show corresponding mass tracks, masses are marked in
  M$_\odot$.}
\end{center}
\end{figure}

To verify that the sources for which we have optical spectra are
indeed members of the Cygnus X complex, we plot the spectroscopic
sample on a dereddened color-magnitude diagram. We used the catalog of
spectral types and intrinsic colors presented in Kenyon and Hartmann
(1995) to determine each star's intrinsic $r'-i'$ color via the Sloan
filter conversions presented in \citet{jordi06}. These colors allow us
to calculate the visual extinction, $A_V$ toward each object using the
relation between $A_\lambda$ and $A_V$ given in
\citet{schlegel98}. The dereddened color-magnitude diagram of the 536
sources in Cygnus X for which we have optical spectra is shown in
Figure~\ref{fig-cmd2} with isochrones from \citet{siess00} shifted to
a distance of 1.7~kpc (red solid lines). We can see that in general
the sources fall between the isochrones at 1 and 5~Myr, which we would
expect given the large amount of embedded star formation in this part
of Cygnus X. In particular, the location of the B stars (shown with
blue stars in Fig.~\ref{fig-cmd2}) is consistent with their being at
the accepted distance of Cygnus X of 1.7~kpc and thus being members of
the region. It is clear however that there are many later type sources
which are above the 1~Myr isochrone, which could be very young stars
or objects in the foreground.

\subsection{Spatial Distribution of YSOs in Cygnus X}

\begin{figure}
\begin{center}
\epsfig{file=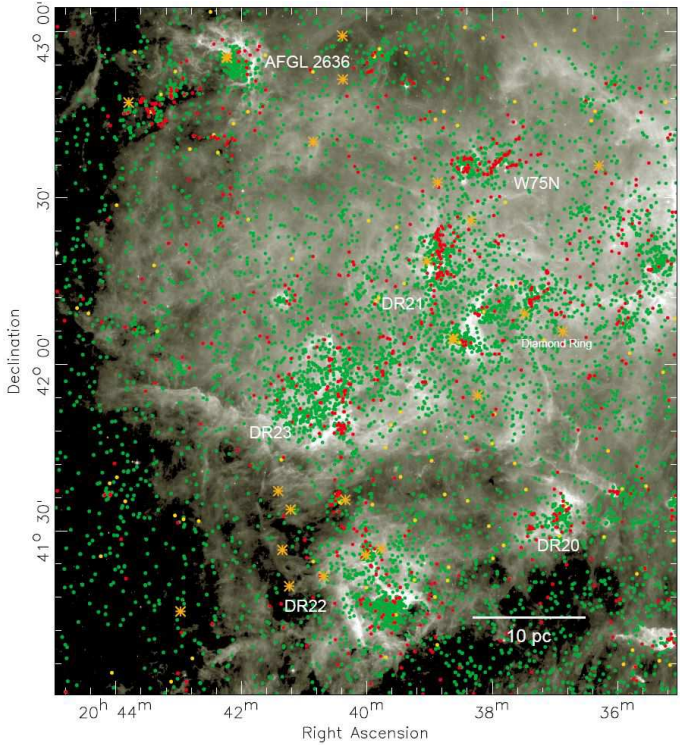,width=1.05\linewidth,clip=}
\caption{Distribution of YSOs overlaid on IRAC channel 4 (8.0 $\mu$m)
  gray-scale image.\label{fig-bigo} \emph{Red}: Class 0 and I;
  \emph{green}: Class II; \emph{yellow}: Transition Disks. Orange
  asterisks indicate B stars. The coordinate axes are in J2000.0
  epoch.}
\end{center}
\end{figure}

Figure~\ref{fig-bigo} shows the spatial distribution of the YSOs
overlaid on the IRAC 8.0~$\mu$m gray-scale image of the north-eastern
part of Cygnus X. The sources are colored according to their
evolutionary class. Although young stars are distributed throughout
the field of view, the figure shows obvious regions of higher stellar
densities. B stars are marked with orange asterisks. The image shows
that the dense clumps of YSOs tend to be concentrated near the massive
B stars. However, there are some B stars that appear distant from any
clustering. Since Cygnus X lies along the Galactic Plane, our source
list of YSOs is likely contaminated by some foreground and background
young stellar populations. We estimate an upper limit to this
contamination by measuring the source density in regions of the lowest
density of YSOs in the complete Cygnus X survey \citep{hora10} that
lie along the Galactic Plane. This method will clearly produce an
overestimate of the level of contamination as many of these sources
will belong to Cygnus X. We find a density of 537 sources per square
degree. We consider this level of contamination low enough to ignore
for the purposes of this particular study.

Many Class I and II objects are clustered together within clouds of
bright IR emission, while others form filamentary structures. For
example, the DR22 region shows a dense cluster of Class II sources
inside a large cavity where the gas appears to have been blown out,
presumably due to the presence of a massive star. This cluster was
also noted by \citet{dutra01} and \citet{leduigou02}. In this region,
we can see Class I objects residing in more filamentary structures
along the rim of the cloud. One branch of Class I objects extends
north in the direction of the DR23 region, along the DR22-DR23
molecular CO filament detected by \citet{schneider06}. DR23 is another
example of a dense concentration of Class II sources surrounded by a
ring of Class I sources along a bright-rimmed cloud. Northwest of this
structure is the \ion{H}{2} region, DR21 \citep{downes66}. The dense
cluster of YSOs in DR21 form an extraordinarily straight chain.

Cygnus X provides a rich and varied collection of YSO clusters for our
study of star formation. To understand the intricate substructure of
the Cygnus X region, we selected two smaller areas in this field of
view to study in more detail. The first region, shown in
Figure~\ref{fig-diamond}, lies just slightly southwest of DR21 and is
termed the Diamond Ring. We can see a bright ring of emission, around
which lie many dense clusters of YSOs, in addition to six identified B
stars. The second region, AFGL~2636, is in the northeast corner of
Figure~\ref{fig-bigo}. Inside an arch of bright emission, there is a
dense group of Class II objects and three B stars. A chain of Class I
sources along the rim of the cloud extends in a long ring to another
dense cluster to the east.

\subsection{Clustered Star Formation in Cygnus X}

To understand the evolution of star formation in Cygnus X, we would
like to be able to identify distinct clusters of young stars and draw
conclusions on their evolution based on comparisons of their relative
sizes and ages. However, there are many obstacles to being able to
glean such information. First, we have no insight on the gravitational
relationships between the stars. Therefore, we must determine cluster
memberships based on the spatial distribution of the sources. Although
we assume all the sources are at about the same distance (1.7~kpc) due
to their IR excess emission, we do not know the exact
three-dimensional position of each star within the cloud. Second,
stars do not form neatly in distinct, independent clusters. Clusters
vary greatly in size, shape and density and can interact and overlap
with other clusters, which renders defining cluster membership
somewhat arbitrary. Therefore, to quantitatively describe the clusters
in Cygnus X in such a way that we can compare with similar studies of
other star-forming regions, we chose to use the cluster isolation
method described and extensively tested in \citet{gutermuth09}. This
method, which relies on the minimal spanning tree (MST) construction
\citep{cartwright04}, has been used recently by Koenig et al. (2008)
to characterize clusters with IRAC and MIPS photometry in W5. Since
our cluster isolation technique relies only on a measure of the local
surface density of stars, it is a $\emph{relative}$ method. Given the
lack of information on the dynamical properties and masses of the
stars, an algorithm to isolate contiguous, locally overdense
structures is the most appropriate tool that we can use in this case.

\subsubsection{The Diamond Ring}

\begin{figure}
\begin{center}
\includegraphics[scale=0.45]{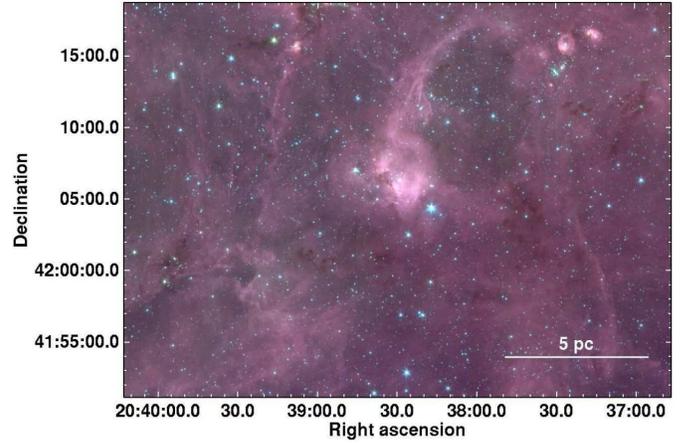}
\caption{IRAC three-color composite image of the Diamond
  Ring.\label{fig-diamond} \emph{red}: 8.0 $\mu$m; \emph{green}: 4.5
  $\mu$m; \emph{blue}: 3.6 $\mu$m.}
\end{center}
\end{figure}

\begin{figure}
\epsscale{2.2}
\plottwo{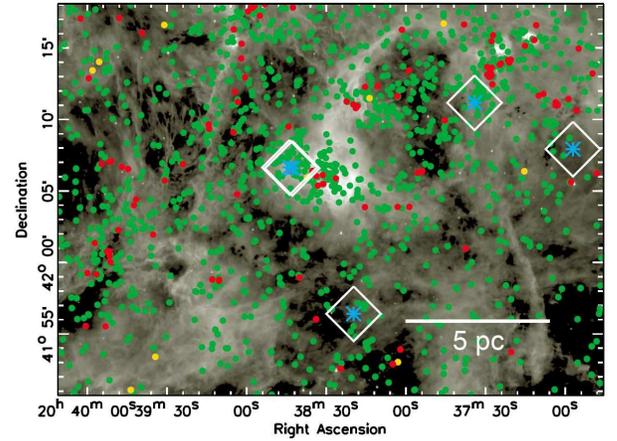}{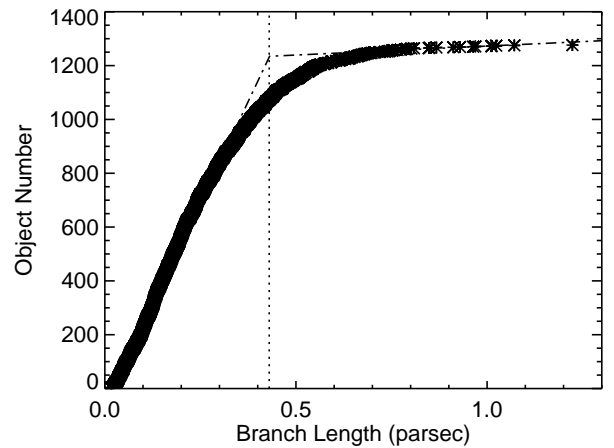} 
  \caption{\emph{Top}: YSO distribution overlaid on IRAC 8.0~$\mu$m
    gray-scale image of the Diamond Ring. \emph{Red}: Class0/ I;
    \emph{green}: Class II; \emph{yellow}: Transition Disks. Blue
    asterisks and white boxes mark B stars. \emph{Bottom}: MST branch
    length distribution.\label{fig-mst-len1} The objects are sorted in
    order of increasing branch length. We fit straight lines through
    the long and short branch length domains. The point of
    intersection, \emph{d$_c$} = 0.43 pc, is chosen for the cutoff
    distance for cluster determination. }
\end{figure}

Located southwest of DR21 lies the region dubbed the $``$Diamond
Ring'' by \citet{marston04}. The Diamond Ring, which is shown in
Figure~\ref{fig-diamond}, is believed to have been formed by a
\ion{H}{2} region identified by a 3 cm radio survey by
\citet{lockman89}. Marston et al. point out that the filaments
extending from DR21, which can be seen in the top-left of
Figure~\ref{fig-diamond}, are truncated by the Diamond Ring and
therefore, conclude that this structure is at the same distance as
DR21. The very bright tip of the Diamond Ring has been identified
previously as the site of a stellar cluster using 2MASS $J$, $H$ and
$K_S$ data (Object 16 and Cl 13 in \citet{dutra01} and
\citet{leduigou02}, respectively). Using the 2MASS photometry and
assuming a distance modulus of 11$^m$, \citet{leduigou02} inferred
that this cluster contains 12 B stars and no O stars. We found two B
stars in the bright tip of the diamond, a B0.5 and a B9.0.  We
identified five B stars in the entire region, which are shown as blue
asterisks in Figure~\ref{fig-mst-len1} (left panel).

\begin{figure}
\epsscale{2.0}
\plottwo{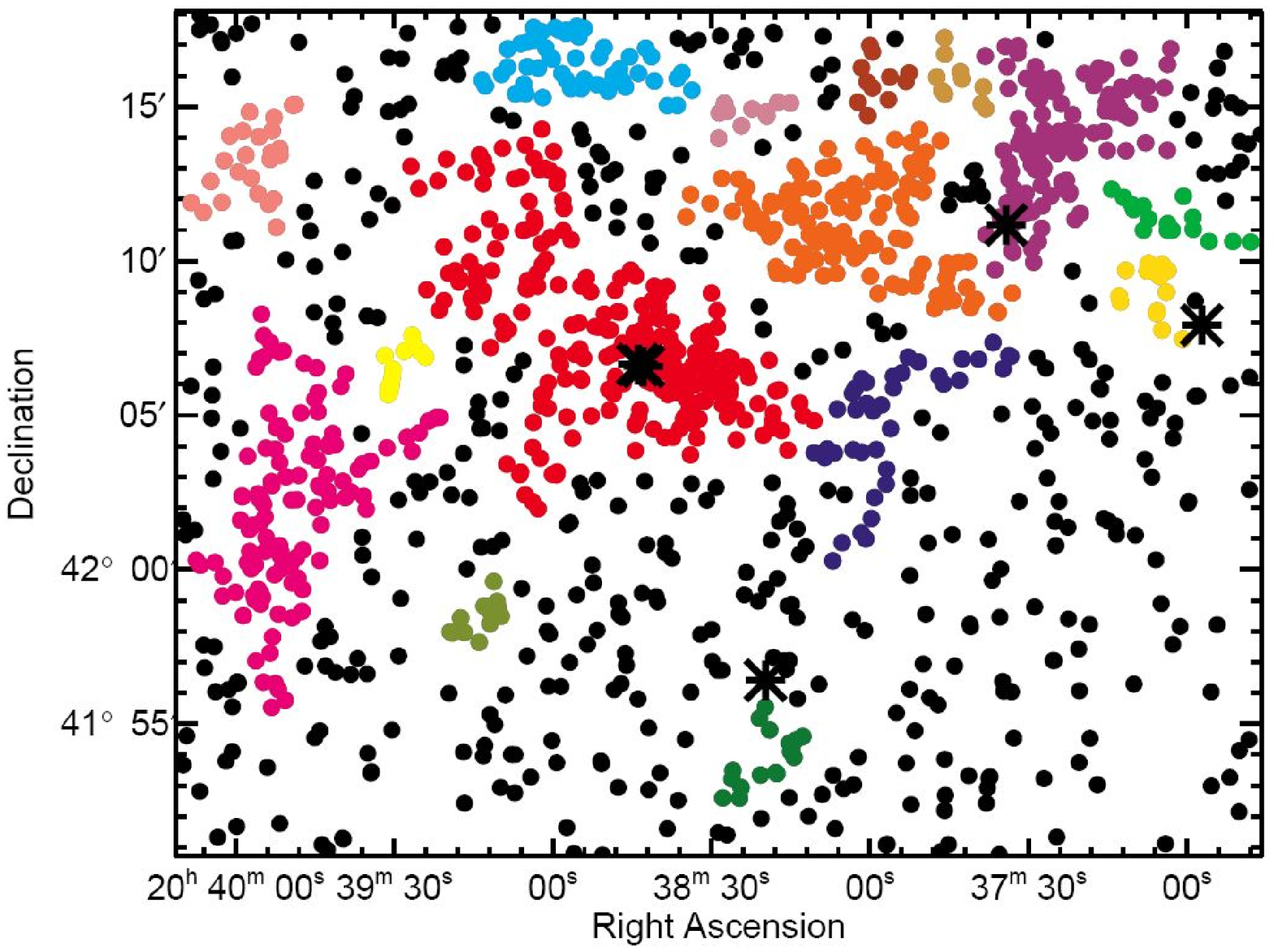}{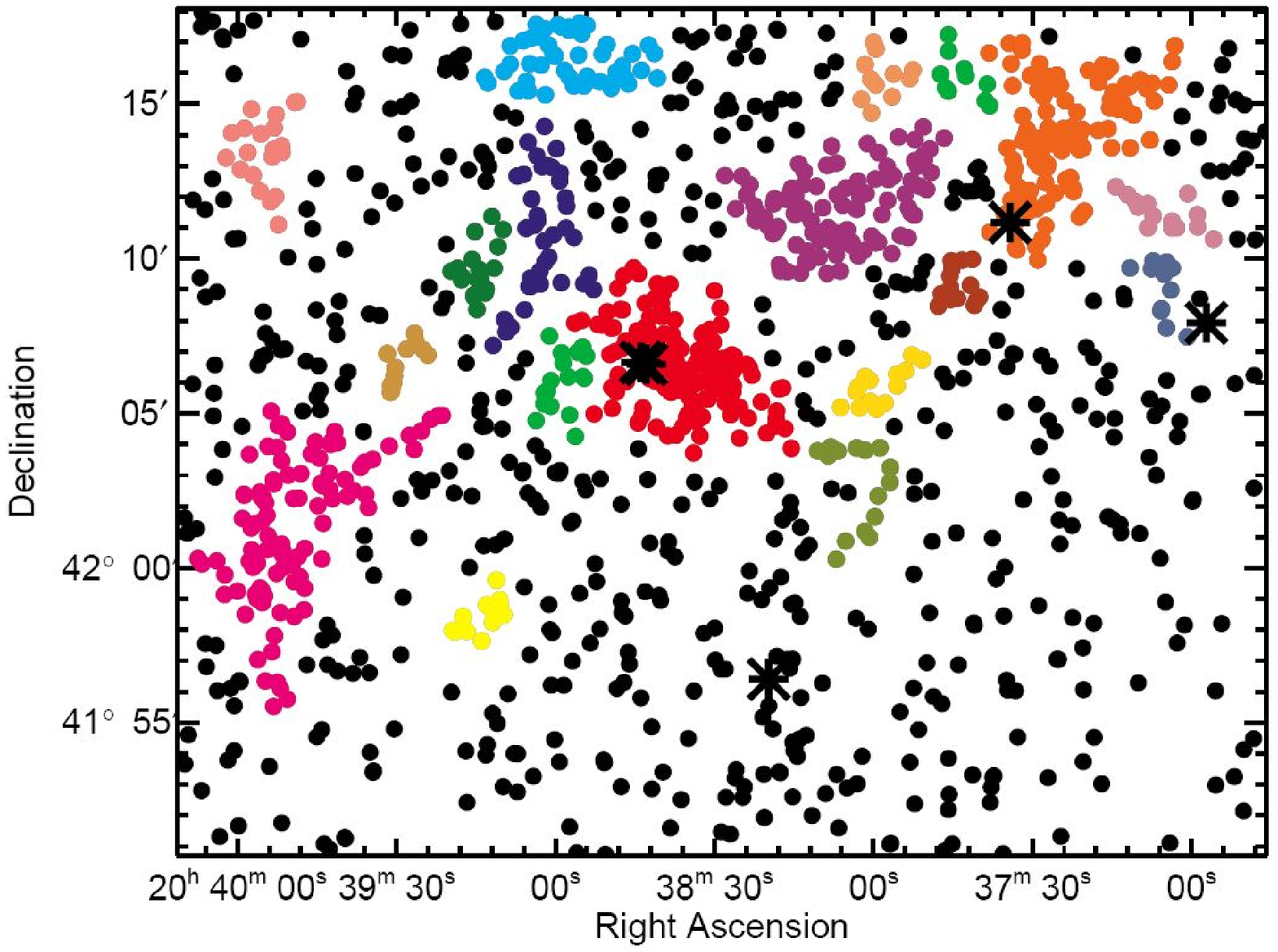}
  \caption{\emph{Top}: Clusters identified using the straight-line
    fit method (\emph{d$_c$}=0.43 pc) are plotted in different
    colors.\label{fig-diam2} Each color represents another cluster and
    the black dots represent stars that are not associated with any
    cluster; \emph{Bottom}: Clusters identified using a cutoff distance
    of \emph{d$_c$} = 0.40 pc are plotted in color, while sources not
    associated with any clusters are plotted in black. Black asterisks
    mark the locations of B stars found in this region.}
\end{figure}

\begin{figure}
\begin{center}
\epsfig{file=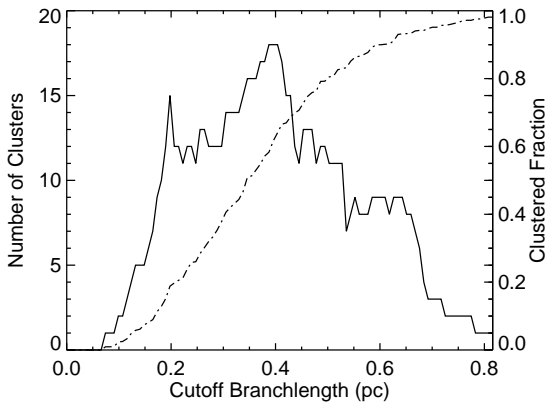,width=0.9\linewidth,clip=}
\caption{\emph{Solid line and left y-scale}: Number of groups
  identified by the MST method;\label{fig-vary1} \emph{Dashed line and
    right y-scale}: Clustered fraction of stars as a function of
  cutoff length.}
\end{center}
\end{figure}

We applied the \citet{gutermuth09} algorithm to identify and
characterize clusters in the Diamond Ring. For each source in the
region, the MST method first finds the distance to its nearest
neighbor. Every source is connected by a line to another source in
such a way that the total length of all the lines is minimized and
there are no closed loops. The distance to its nearest neighbor is
called the star's branch length. Clusters are defined as a collection
of stars that are connected by branches that are less than some cutoff
distance and containing at least some minimum number of stars, N. We
define the minimum number of stars to be N=10, attempting to minimize
the identification of false groupings due to statistical fluctuations
in the low density YSO distribution. To determine the cutoff distance,
we plot the distribution of branch lengths in
Figure~\ref{fig-mst-len1} (right). The distribution shows two distinct
regions---a short and a long branch length domain---corresponding to
clustered and distributed sources respectively. While there is no
definitive cutoff branch length that perfectly determines cluster
membership, choosing a cutoff length that falls somewhere between
these two regions is a good estimate of the divide between objects
belonging to clusters and the distributed population. We fit lines
through the short and long branch length parts of the distribution and
extrapolate to find their intersection. The point of intersection is
selected to be the cutoff length, \emph{d$_c$}, which here equals
0.43~pc. Thus, we defined clusters in the Diamond Ring region as any
group of 10 or more stars connected by lengths of no more than
0.43~pc. This cutoff distance is about half that derived by Koenig et
al. (0.86 pc) for W5. This shorter cutoff length suggests that the
Diamond Ring may be more densely populated with YSOs than W5.

\begin{deluxetable}{lcc}
\tablecaption{Diamond Ring Cluster Summary\tablenotemark{a}\label{tbl-3}}
\tablewidth{0pt}
\tablehead{\colhead{Parameter} & \colhead{Straight-Line Fit} & \colhead{Ngrp Max} } 
\startdata
Number of Clusters    &     15    &  18 \\
Percent in Clusters   &     67.16 &  58.31 \\
Total Number in Clusters & 857    & 744 \\
Group Size      &      10-255     & 10-148\\
Cutoff Distance &      0.43~pc  & 0.4~pc
\enddata
\tablenotetext{a}{Summary of clusters identified by the cluster isolation
   algorithm using the straight-line fit method with a cutoff distance
  \emph{d$_c$} = 0.43 pc and using \emph{d$_c$} = 0.40 pc to maximize
  the number of clusters.}
\end{deluxetable}

\begin{deluxetable*}{lccccccccc}
\tablecaption{Diamond Ring Clusters\label{tbl-5}}
\tabletypesize{\small}
\tablewidth{0pt}
\tablehead{\colhead{Cluster} & \colhead{R.A. (J2000.0)} & \colhead{Dec. (J2000.0)} & \colhead{N$_{IR}$\tablenotemark{a}} & \colhead{I} & \colhead{II} & \colhead{II/I\tablenotemark{b}} & \colhead{N$_{emb}$\tablenotemark{c}} & \colhead{N$_{td}$\tablenotemark{d}} & \colhead{Diameter}\\
\colhead{} & \colhead{(h m s)} & \colhead{($\degr$ $\arcmin$ $\arcsec$)} & \colhead{}  & \colhead{}  & \colhead{}  & \colhead{}  & \colhead{} & \colhead{} & \colhead{(pc)} }
\startdata
G81.51+0.43 & 20:38:50.06 & +42:07:55.83 & 255 & 14 & 237 & 16.9(4.7) & 3 & 1 & 8.1\\
G81.48+0.61 & 20:38:06.27 & +42:11:48.22 & 133 & 10 & 122 & 12.2(4.0) & 0 & 1 & 5.8 \\
G81.44+0.75 & 20:37:26.73 & +42:14:38.67 & 132 & 22 & 110 & 5.0(1.2) & 0 & 0 & 4.7\\
G81.37+0.55 & 20:37:57.51 & +42:05:25.25 & 39 & 4 & 35 & 8.8(4.6) & 0 & 0 & 4.5\\
G81.71+0.34 & 20:39:59.88 & +42:13:18.00 & 24 & 1 & 21 & 21.0(21.5) & 0 & 2 & 2.4\\
G81.27+0.39 & 20:38:19.82 & +41:54:25.17 & 17 & 1 & 16 & 16.0(16.5) & 0 & 0 & 1.7\\
G81.35+0.76 & 20:37:05.44 & +42:12:11.80 & 16 & 7 & 9 & 1.3(0.6) & 0 & 0 & 2.6\\
G81.42+0.30 & 20:39:14.40 & +41:58:47.48 & 14 & 3 & 11 & 3.7(2.4) & 0 & 0 & 1.1\\
G81.56+0.35 & 20:39:31.17 & +42:06:43.78 & 13 & 2 & 11 & 5.5(4.2) & 0 & 0 & 1.0\\
G81.35+0.72 & 20:37:08.45 & +42:09:48.97 & 13 & 0 & 13 & \nodata & 0 & 0 & 1.5\\
G81.53+0.66 & 20:38:01.26 & +42:16:29.25 & 11 & 1 & 10 & 10.0(10.5) & 0 & 0 & 1.1\\
G81.54+0.61 & 20:38:25.87 & +42:15:18.75 & 10 & 0 & 10 & \nodata & 0 & 0 & 1.4\\
G81.51+0.69 & 20:37:46.00 & +42:16:32.79 & 10 & 0 & 9 & \nodata & 0 & 1 & 1.3
\enddata
\tablenotetext{a}{Number of stars with infrared excess. Includes Class I, II, deeply embedded protostars and transition disk candidates.} 
\tablenotetext{b}{Number in parentheses indicates Poisson uncertainty in ratio.} 
\tablenotetext{c}{Number of deeply embedded protostars.} 
\tablenotetext{d}{Number of transition disk candidates.} 
\end{deluxetable*}

\begin{figure}
\epsscale{1.1}
\plotone{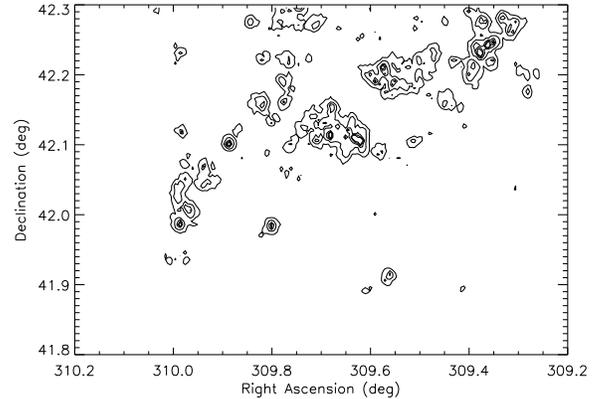}
\caption{Stellar surface density contour plot for the Diamond
  Ring.\label{fig-contour1} Contours are drawn at 20, 40, 60, and 80
  $pc^{-2}$}
\end{figure}

With these criteria we identified 15 clusters, which contain 67~\% of
the sources. In Figure~\ref{fig-diam2} (left panel), the clusters are
shown in color with the distributed sources in black. B type stars are
marked with black asterisks. The clusters exhibit a wide spread in
size, ranging from 10 to 255 members. Most of the larger clusters are
fairly round, while some of the smaller ones are more
filamentary. While most of the B type stars appear embedded inside
clusters, there is one that lies just outside a smaller, more
filamentary group to the South. This result is similar to what was
seen in W5. While most massive stars were within large clusters,
Koenig et al. found one O type star that did not lie within any
cluster. However, in both W5 and the Diamond Ring, these massive stars
are still found in relatively dense areas.

To investigate how our result depends on the cutoff length, we
searched for the maximum number of clusters (of 10 or more stars) that
could be identified by varying the cutoff length. We identified
clusters for cutoff lengths from 1 to 100$\arcsec$, with length steps
of 1$\arcsec$. Figure~\ref{fig-vary1} shows the number of clusters and
the clustered fraction as a function of cutoff distance. As
\emph{d$_c$} increases, more and more groups meet the cluster
criteria. The number of clusters reaches a maximum at 18 and then
begins to decline. As the cutoff length continues to increase, groups
of smaller clusters get classified as one large cluster. Thus,
choosing a good value for \emph{d$_c$} relies on a compromise between
including the less dense clusters and not over-simplifying
substructure in the complex regions. The technique of maximizing the
number of groups to determine the cutoff length was investigated in
detail by \citet{batti91}.

A maximum of 18 clusters were identified for \emph{d$_c$} =
0.40~pc. The new clusters (shown in Fig.~\ref{fig-diam2}, right) look
similar to those for \emph{d$_c$} = 0.43~pc. Two of the largest
clusters were split into smaller groups, one was not found at all, and
a few stars became unassociated with any cluster. An increase from 15
to 18 clusters implies that the original 0.43 pc is a reasonable
cutoff distance for identifying clusters. Table~\ref{tbl-3} summarizes
the cluster characteristics for each method. The fact that the overall
cluster fraction appears to have changed relatively little from one
value of \emph{d$_c$} to the next provides confidence that this method
is objective enough for identifying clusters in the Diamond Ring.

Table~\ref{tbl-5} lists the properties of each cluster in the Diamond
Ring as derived from the straight-line fit method with $d_c$ =
0.43~pc. We omit two clusters that are part of much larger clusters
outside this region (the large pink cluster centered roughly at
20$^h$39$^m$45$^s$, 42$\degr$02$\arcmin$ and the light blue cluster at
20$^h$9$^m$, 42$\degr$16$\arcmin$, Fig.~\ref{fig-diam2}, left). The
clusters are named by the Galactic coordinates of the cluster centers,
as determined by averaging the positions of all members. For each of
the 13 listed clusters in the region, we give the number of stars with
IR excess, which includes Class I and II sources and any deeply
embedded protostars and transition disk sources. We list the number of
each individual class and the ratio of Class II to Class I sources for
each cluster. We note that small-number statistics make the ratios for
the smaller clusters very uncertain: we assume Poisson uncertainty
applies and list this value in Table~\ref{tbl-5} alongside each quoted
ratio. The II/I ratios range from 1.3$\pm$0.6 to 21.0$\pm$21.5. The
final column in the table gives the diameter of each cluster in
parsecs, measured as the diameter of the smallest circle that
encompasses all sources in that cluster.

We constructed a 9$\arcsec$ grid in the Diamond Ring region and at
each point calculated the stellar surface density given by:

\begin{equation}
\sigma=\frac{N-1}{\pi r_N^2}
\end{equation}

\noindent where $r_N$ is the distance to the Nth nearest star and N=5
\citep{casertano85}. The surface density contour map for the Diamond
Ring is shown in Figure~\ref{fig-contour1}. The major groups
identified in this density plot agree well with the clusters
identified with the MST/cluster isolation algorithm.

\subsubsection{AFGL~2636}

\begin{figure}
\begin{center}
\includegraphics[scale=0.45]{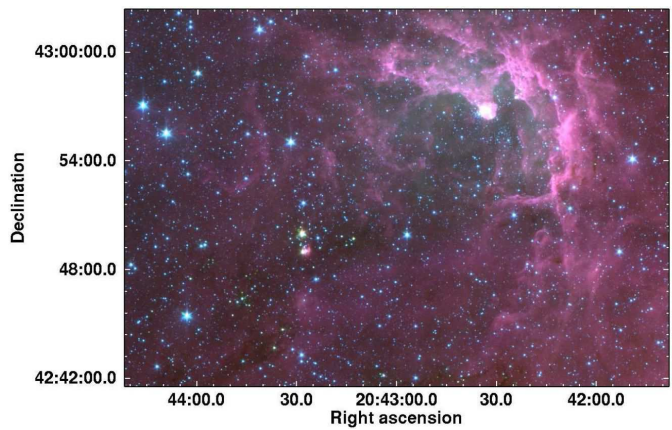}
\caption{IRAC three-color composite image of AFGL
    2636.\label{fig-2636} \emph{red}: 8.0 $\mu$m; \emph{green}: 4.5
    $\mu$m; \emph{blue}: 3.6 $\mu$m.}
\end{center}
\end{figure}

\begin{figure}
\epsscale{2.2}
\plottwo{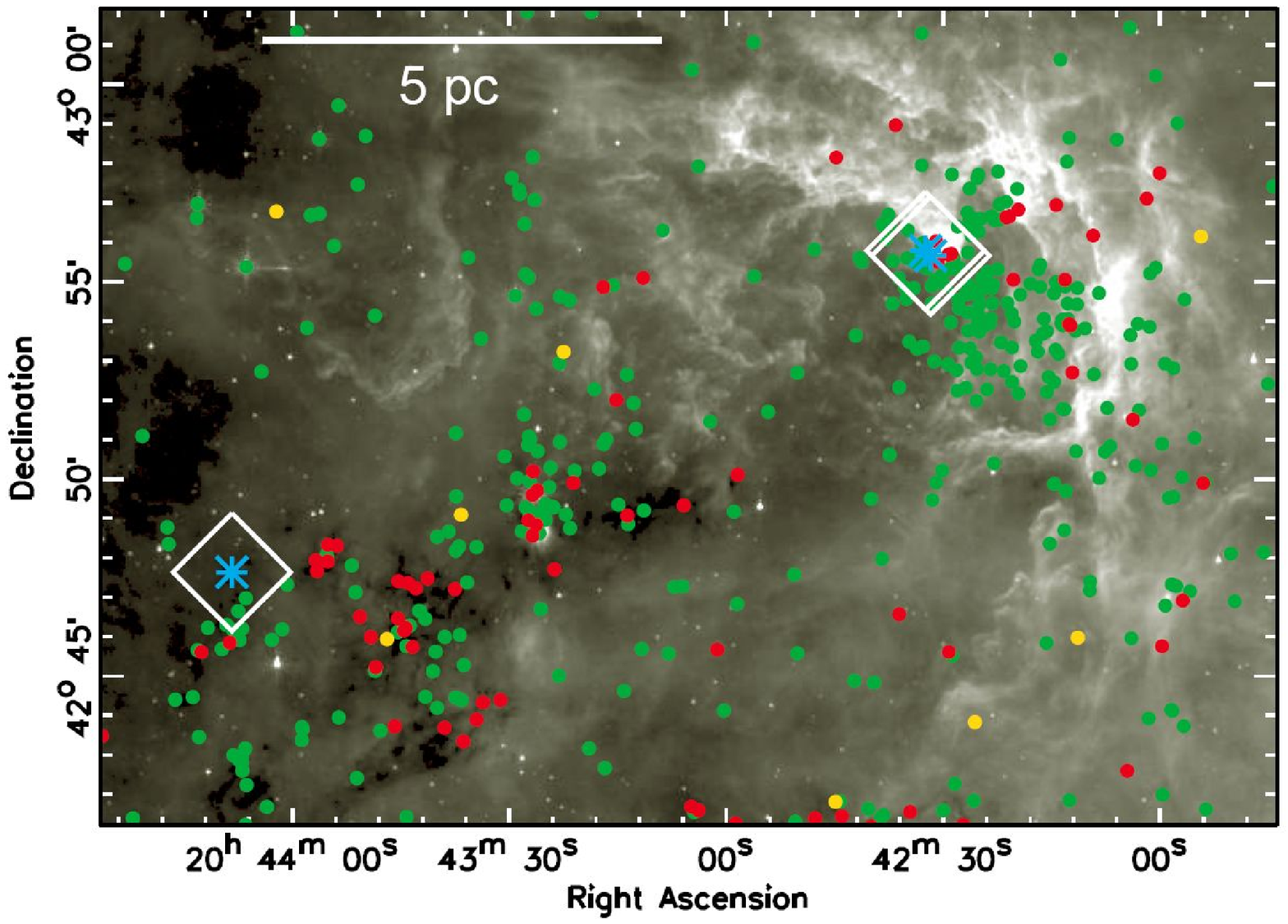}{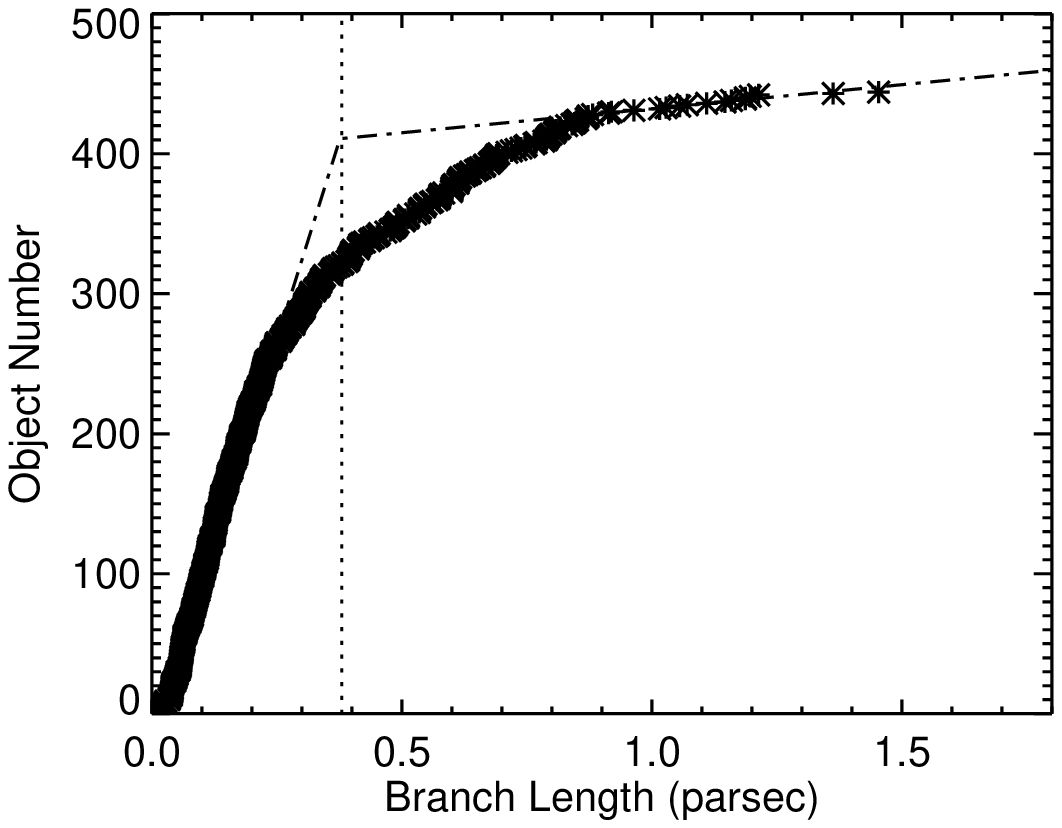}
  \caption{\emph{Top}: YSO distribution overlaid on IRAC channel 4 gray-scale (8.0
    $\mu$m) image of AFGL~2636.\label{fig-reg2-overlay} \emph{Red}:
    Class0/ I; \emph{green}: Class II; \emph{yellow}: Transition
    Disks. Blue asterisks and white boxes mark B stars. \emph{Bottom}:
    MST branch length distribution for AFGL~2636 region. We fit
    straight lines through the long and short branch length
    domains. The point of intersection, \emph{d$_c$} = 0.38 pc, is
    chosen for the cutoff distance for cluster determination.}
\end{figure}

The second region studied, which encompasses AFGL~2636, lies in the
northeast corner of Cygnus X North. Figure~\ref{fig-2636} shows an
IRAC channel 1, 2 and 4 color composite image of the
region. Figure~\ref{fig-reg2-overlay} displays the spatial
distribution of the YSOs on a gray-scale IRAC channel 4 image. Two B
stars, a B0.0$\pm$2.0 and a B5.0$\pm$3.0, sit very close together at
the center of the bright molecular cloud. The B0.0 star was first
classified by \citet{gehrz80} as a young B1-2 source. A dense cluster
of Class II objects is distributed around the B stars and lies nestled
within the bright-rimmed cloud. This cluster was first identified by
\citet{dutra01}, who associated it with the radio \ion{H}{2} region
G82.6+0.4. A string of Class I sources lies along the rim of the cloud
and extends in a wide arch to another dense group of YSOs, which
reside in a dark region in the left of the image.

\begin{figure}
\epsscale{2.0}
\plottwo{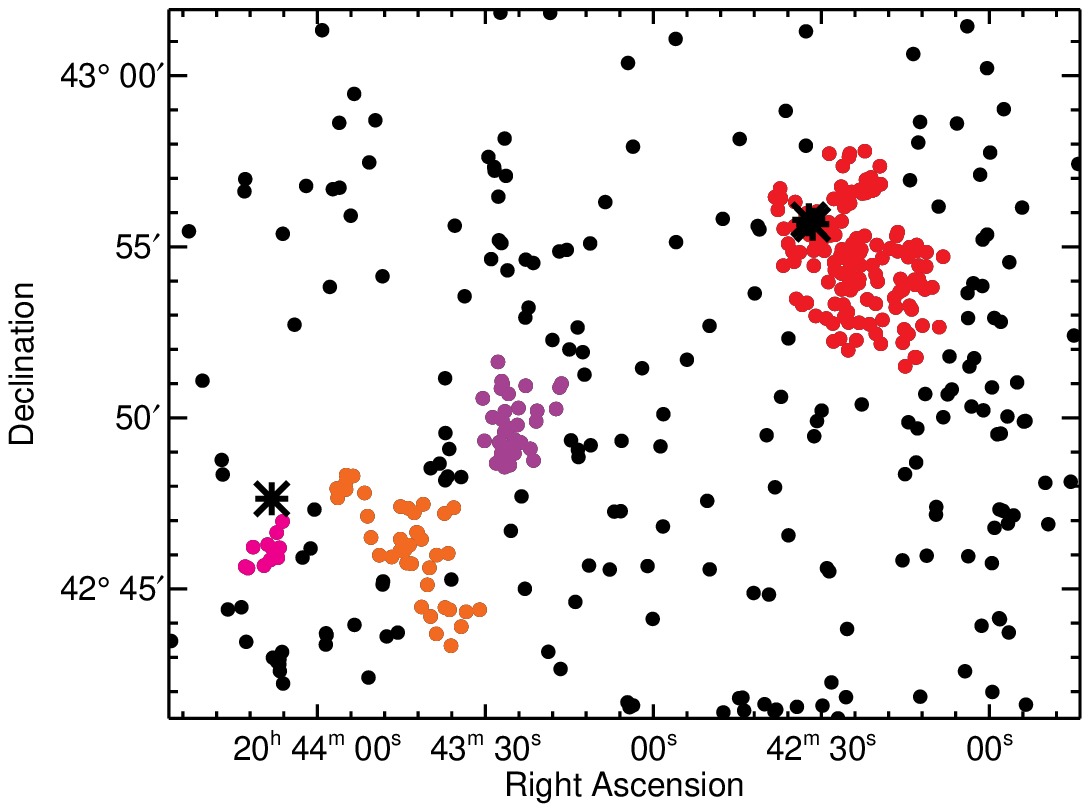}{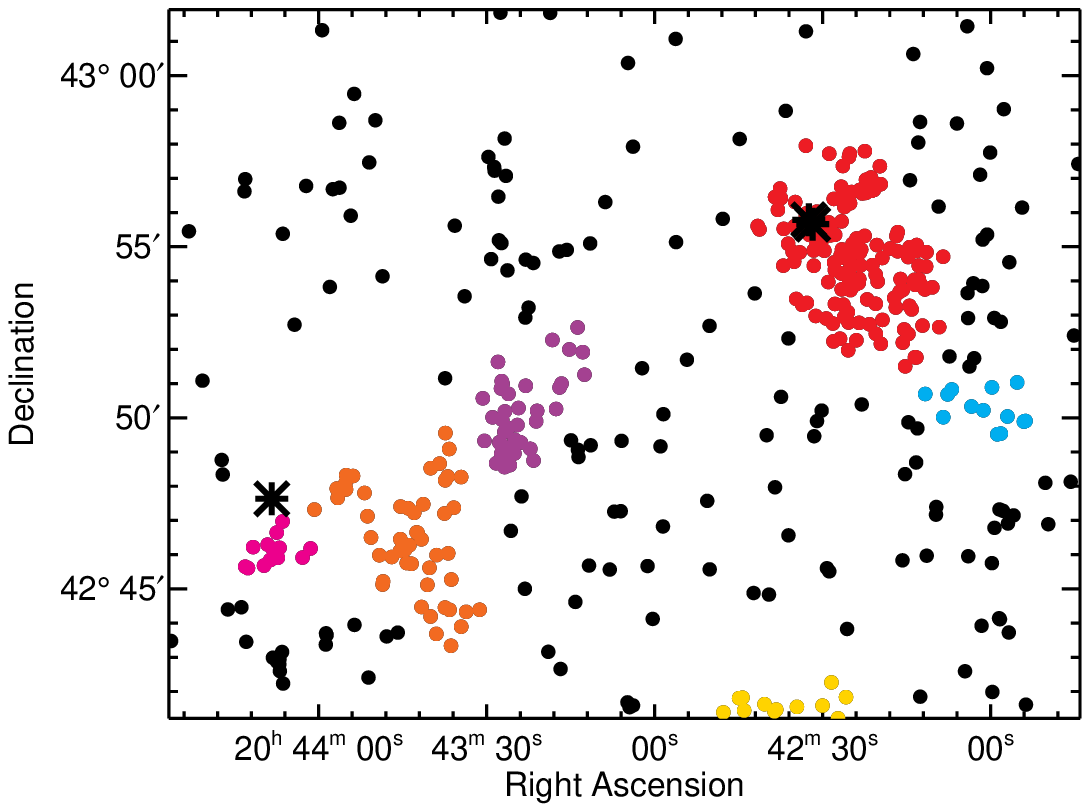}
  \caption{\emph{Top}: Clusters identified using the straight-line
    fit method (\emph{d$_c$} = 0.38~pc).\label{fig-2636b} Each color
    represents another cluster and the black dots represent stars that
    are not associated with any cluster; \emph{Bottom}: Clusters
    identified using a cutoff distance of 0.42~pc are plotted in
    color, while sources not associated with any clusters are plotted
    in black.}
\end{figure}

\begin{figure}
\plotone{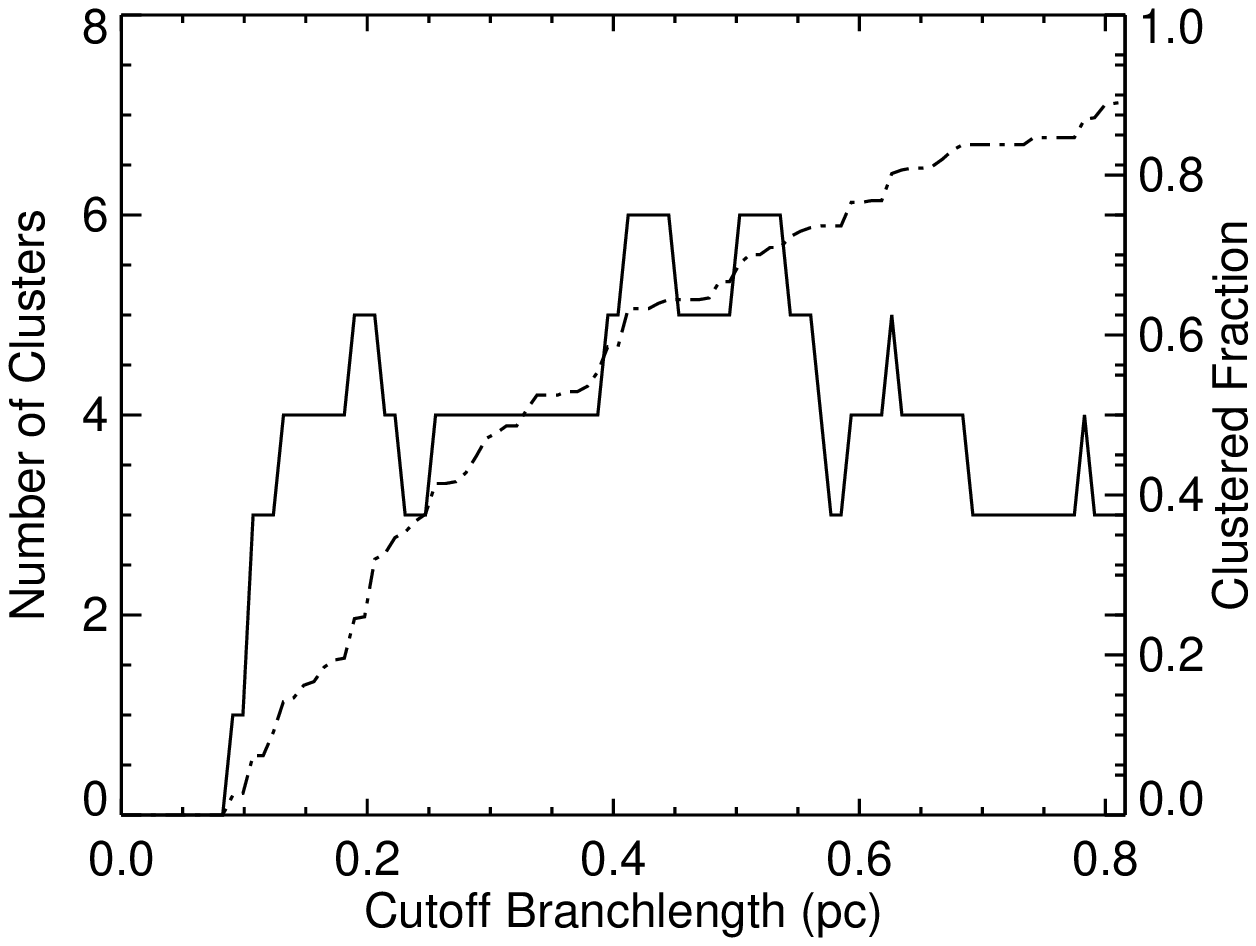}
  \caption{\emph{Solid line and left y-scale}:\label{fig-vary2} Number
    of groups identified by the MST method; \emph{Dashed line and
      right y-scale}: Clustered fraction of stars as a function of
    cutoff length.}
\end{figure}

\begin{deluxetable}{lcc}
\tablecaption{AFGL~2636 Cluster Summary\label{tbl-4}}
\tablewidth{0pt}
\tablehead{\colhead{Parameter} & \colhead{Straight-Line Fit} & \colhead{Ngrp Max}}
\startdata
Number of Clusters & 4 & 6 \\
Percent in Clusters & 52.90 & 63.29 \\
Total Number in Clusters & 235 & 281 \\
Group Size & 11-146 & 12-149\\
Cutoff Distance & 0.38~pc  & 0.42~pc \\
\enddata
\end{deluxetable}

We performed the same clustering isolation algorithm to this
region. The distribution of branch lengths in the MST is shown in
Figure~\ref{fig-reg2-overlay} (right panel). The straight-line fit
yielded a cutoff distance of \emph{d$_c$} = 0.38~pc, which led to the
identification of four clusters with a clustered fraction of 52.9
percent. The clusters, which are plotted in colors in the left panel
of Figure~\ref{fig-2636b}, range in size from 11 to 146 members. This
cutoff distance is comparable to that used in the Diamond Ring region,
\emph{d$_c$} = 0.43~pc. The fact that the cutoff distance determined
by the straight-line fit for AFGL~2636 is slightly shorter than for
the Diamond Ring implies that the clusters in AFGL~2636 are denser.

As with the Diamond Ring, we varied the cutoff length in the cluster
isolation algorithm from 1 to 100$\arcsec$ in 1$\arcsec$ steps. The
number of clusters and the clustered fraction that resulted for each
cutoff distance is plotted in Figure~\ref{fig-vary2}. The number of
clusters vs. cutoff length plot shows two maxima at six clusters. A
cutoff distance of \emph{d$_c$} = 0.42 pc results in a total of 6
clusters and a larger clustered fraction of 63.3\%. This cutoff
distance identified the same clusters as the straight-line fit method
(\emph{d$_c$} = 0.38~pc), but with the addition of two new
clusters. The two new clusters identified, G82.45+0.42 and
G82.40+0.23, seem to exhibit definite structure, and we therefore
chose to include them in our analysis. Table~\ref{tbl-6} lists the
parameters of each of the six clusters, including the number of stars
with IR excess, the number of each individual class and the ratio of
Class II to Class I. The II/I ratios range from 1.6$\pm$0.5 to
12.5$\pm$3.9. The final column gives the cluster diameters in parsecs
calculated in the same way as for the Diamond Ring clusters.

\begin{figure}
\epsscale{1.1}
\plotone{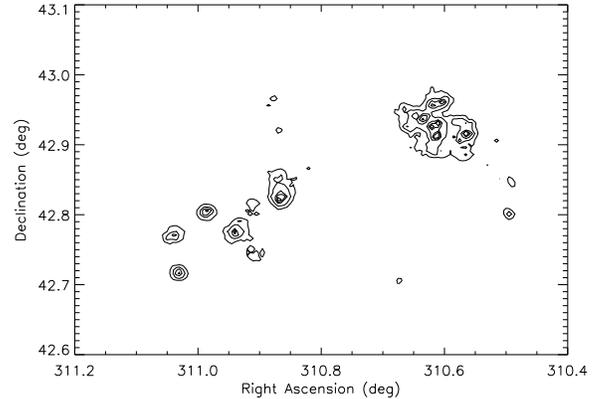}
  \caption{Stellar surface density contour plot of
    AFGL~2636.\label{fig-contour2} Contours are drawn at 20, 40, 60,
    and 80 $pc^{-2}$.}
\end{figure}

We constructed a 9$\arcsec$ grid in AFGL~2636 and at each point
calculated the stellar surface density given by Equation 5. The
surface density contour map for AFGL~2636 is shown in
Figure~\ref{fig-contour2} with contours at 20, 40, 60, and 80
$pc^{-2}$. The major groups identified in this density plot agree
quite well with the clusters identified with the cluster isolation
algorithm.

\begin{deluxetable*}{lccccccccc}
\tablecaption{AFGL~2636 Clusters\label{tbl-6}}
\tabletypesize{\small}
\tablewidth{0pt}
\tablehead{\colhead{Cluster} & \colhead{R.A. (J2000.0)} & \colhead{Decl. (J2000.0)} & \colhead{N$_{IR}$\tablenotemark{a}} & \colhead{I} & \colhead{II} & \colhead{II/I\tablenotemark{b}} & \colhead{N$_{emb}$} & \colhead{N$_{td}$} & \colhead{Diameter}\\
\colhead{} & \colhead{(h m s)} & \colhead{($\degr$ $\arcmin$ $\arcsec$)} & \colhead{}  & \colhead{}  & \colhead{}  & \colhead{}  & \colhead{} & \colhead{} &\colhead{(pc)}}
\startdata
G82.55+0.40 & 20:42:26.83 & +42:55:33.46 & 149 & 11 & 137 & 12.5(3.9) & 1 & 0 & 3.6 \\
G82.57+0.12 & 20:43:44.63 & +42:46:50.87 & 52 & 18 & 28 & 1.6(0.5) & 4 & 2 & 3.1\\
G82.58+0.21 & 20:43:25.75 & +42:50:28.60 & 42 & 6 & 34 & 5.7(2.5) & 2 & 0 & 2.4\\
G82.61+0.06 & 20:44:09.14 & +42:46:14.40 & 13 & 2 & 11 & 5.5(4.2) & 0 & 0 &  1.1\\
G82.45+0.42 & 20:42:02.85 & +42:51:17.71 & 13 & 1 & 12 & 12.0(12.5) & 0 & 0 & 1.7\\
G82.40+0.23 & 20:42:37.11 & +42:42:24.11 & 12 & 4 & 7 & 1.8(1.1) & 0 & 1 & 2.0
\enddata
\tablenotetext{a}{Number of stars with infrared excess. Includes Class I, II, deeply embedded protostars and transition disk candidates.} 
\tablenotetext{b}{Number in parentheses indicates Poisson uncertainty in ratio.} 

\end{deluxetable*}

\section{Discussion}

\subsection{Star Formation in The Diamond Ring}

\begin{figure}
\epsfig{file=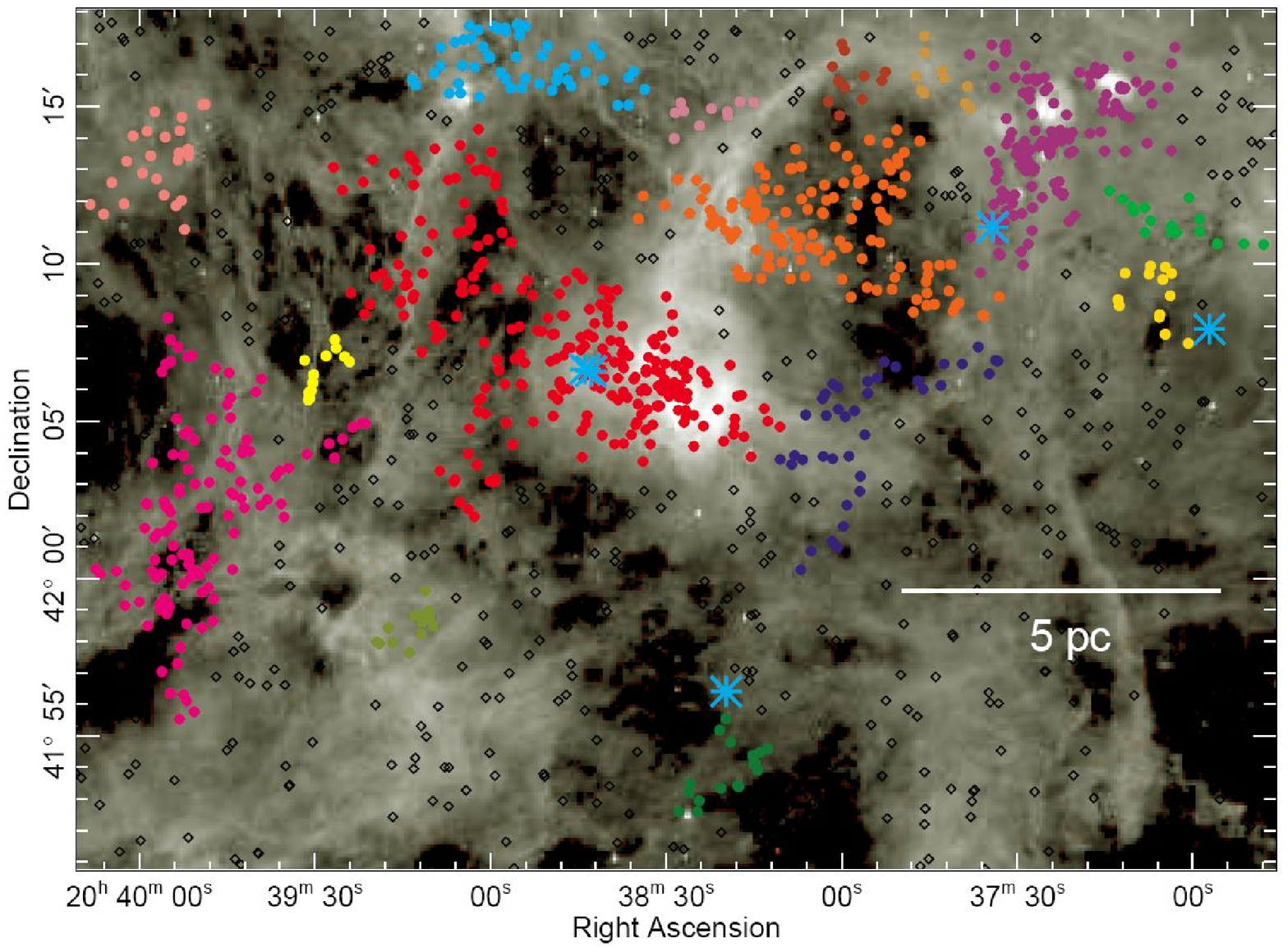,width=0.91\linewidth,clip=}
\epsfig{file=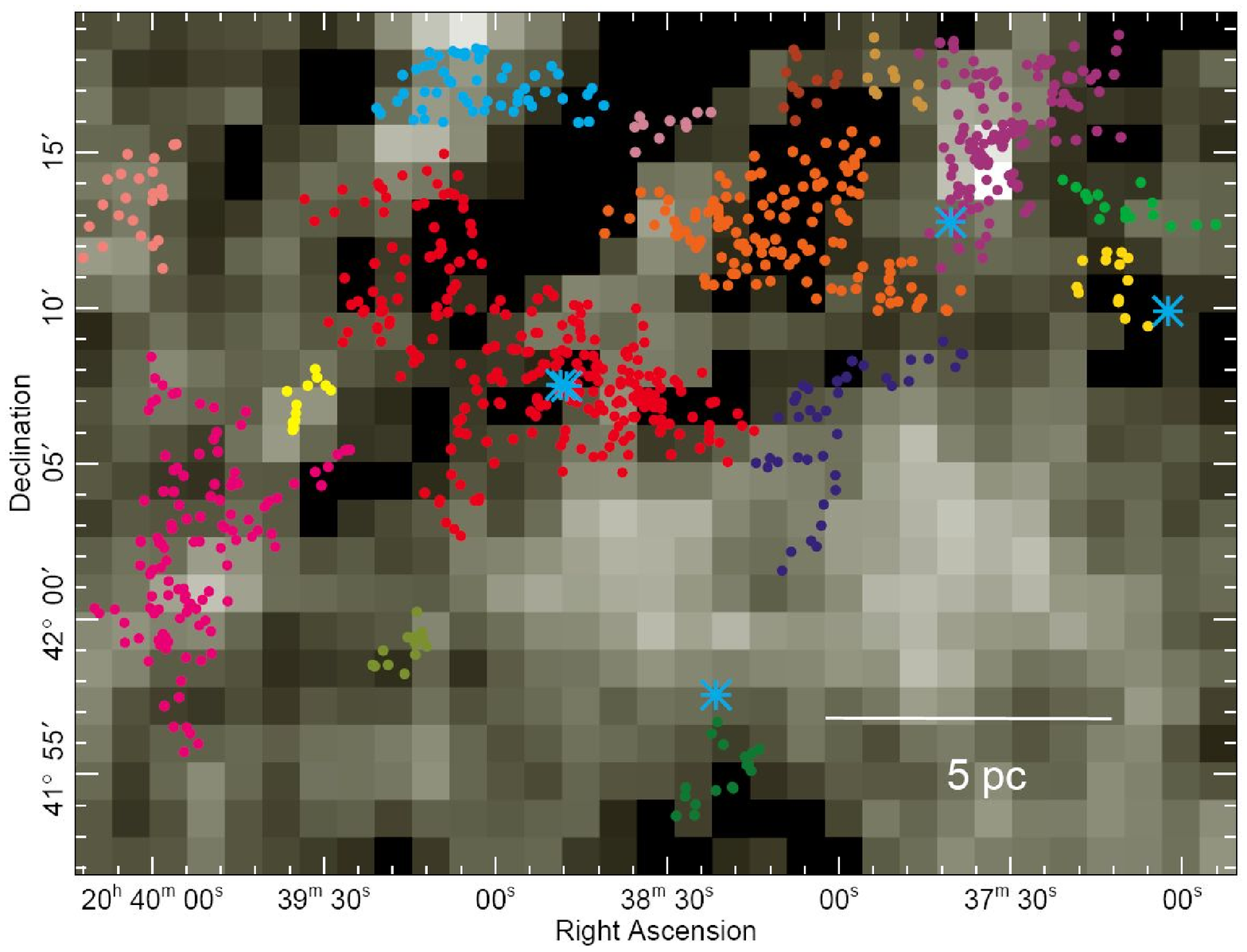,width=0.91\linewidth,clip=}
\caption{\emph{Top}: The clusters identified in the Diamond Ring
  region using \emph{d$_c$} = 0.43~pc overlaid on an IRAC channel 4
  gray-scale image.\label{fig-reg1-over} \emph{Bottom}: $A_V$
  extinction map of the Diamond Ring generated with 2MASS photometry
  with clusters overlaid. White indicates higher extinction. Blue
  asterisks mark B stars.}
\end{figure}

Using the MST/cluster isolation method, we have identified 13 clusters
in the Diamond Ring region, which are shown overlaid on an IRAC
channel 4 image in Figure~\ref{fig-reg1-over} (Left panel). To better
understand the evolution of star-formation in these regions, we next
need to determine the ages of each cluster. Different types of stars
require different amounts of time to progress through the various
evolutionary stages. This variability makes determining the ages of
the clusters difficult, because we cannot say that one Class I object
is definitely younger than another Class II object. However, since
Class I objects do represent an earlier stage of star formation than
Class II, we can use the ratio of II/I sources to estimate the
relative ages of the cluster. Table~\ref{tbl-5} lists the II/I ratio
for each cluster in the Diamond Ring. Clusters with higher II/I ratios
are dominated by Class II sources, and therefore should be relatively
older.

According to the II/I ratios, the red cluster, G81.51+0.43, appears to
be the oldest with a II/I ratio of 16.9$\pm$4.7 (G81.71+0.34 has a
higher ratio but with much larger uncertainty). This cluster, which
sits at the tip of the Diamond Ring and contains three B stars, is
dominated by Class II sources and perhaps represents the earliest
phase of star formation in the Diamond Ring region. One of the oldest
clusters, G81.48+0.61 (orange), lies at the center of the Diamond
Ring. With a II/I ratio of 12.2$\pm$4.0, it may have formed slightly
after the red cluster. We have the optical spectrum from only one star
in this orange cluster, which was found to be an F type star.

Most of the other clusters that surround the red and orange clusters
have lower II/I ratios. G81.44+0.75 (purple) contains 22 class I
sources, which is the most of all the clusters in the Diamond Ring. It
has a II/I ratio of 5.0$\pm$1.2, implying that it is younger than the
red and orange. Other younger clusters include G81.35+0.76 (green, top
right), G81.42+0.30 (olive, bottom left), G81.56+0.35 (yellow, left)
and G81.37+0.55 (blue) with II/I ratios of 1.3$\pm$0.6, 3.7$\pm$2.4,
5.5$\pm$4.2, and 8.8$\pm$4.6, respectively. These clusters lie in a
ring around the older red and orange clusters.

The distribution of these clusters helps us to understand the star
formation history of the Diamond Ring. It appears that the initial
star formation occurred in the red and orange clusters. Then winds and
radiation from the most massive stars in the red cluster sent shocks
through the dust and gas in the surrounding regions, possibly inducing
the subsequent generation of star formation, represented by the
smaller, younger surrounding clusters. This situation is similar to
what was found in W5 and NGC~602 \citep{koenig08,
  carlson07}. Figure~\ref{fig-reg1-over} (right panel) shows the
clusters plotted on the visual extinction map of this region, where
white indicates higher extinction. The red and orange clusters lie in
very dark regions, implying little extinction. The young clusters
(purple, green, blue, olive, yellow), on the contrary, tend to lie in
areas of higher extinction. These young clusters still reside within
their dusty natal molecular clouds. However, in the older red and
orange clusters, winds and radiation from the massive stars have
effectively removed the dust and gas from their parental molecular
clouds. Although we did not find any O type stars in our spectral
survey, several objects coincident with the red and orange clusters
have $r'i'H\alpha$ photometry consistent with a spectral type earlier
than B0 at the distance and extinction of Cygnus X. The fact that the
red and orange clusters have had time to blow away their molecular
clouds while the others have not is consistent with our picture of
star formation in the Diamond Ring.

\subsection{Star Formation in AFGL~2636}

\begin{figure}
\epsfig{file=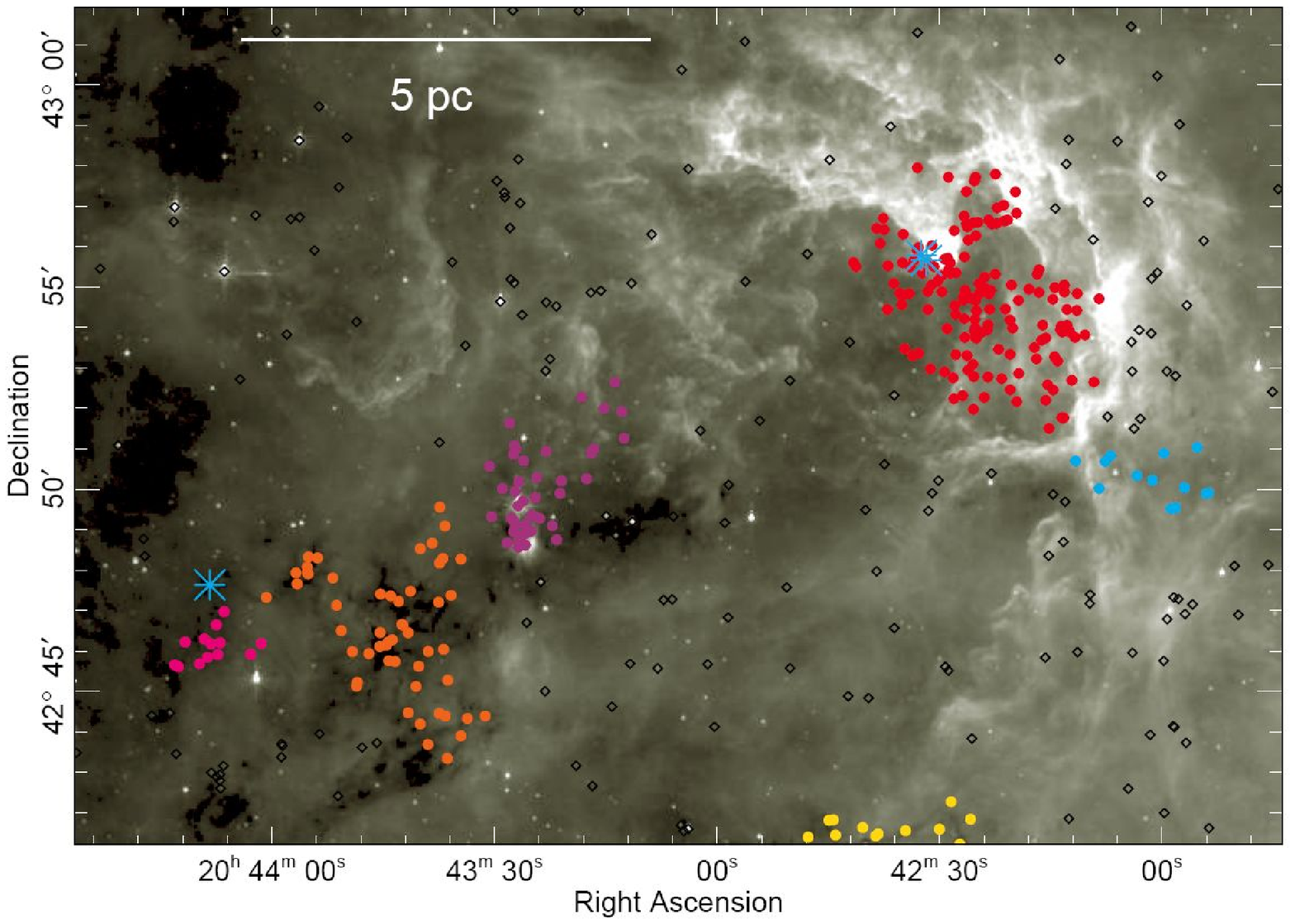,width=0.91\linewidth,clip=}
\epsfig{file=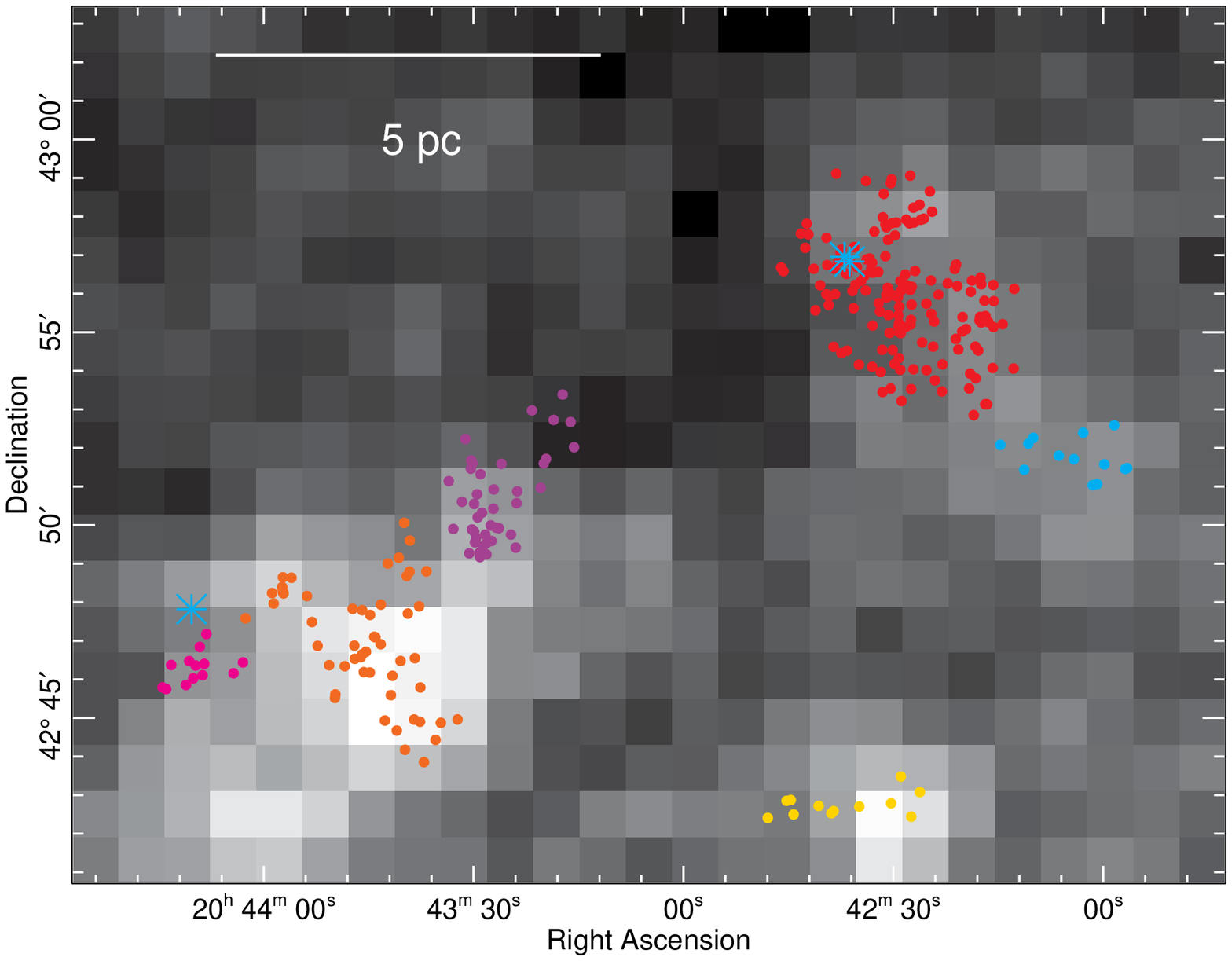,width=0.91\linewidth,clip=}
\caption{\emph{Top}: Clusters identified using \emph{d$_c$} = 0.42~pc
  overlaid on IRAC channel 4 gray-scale image of AFGL~2636;
  \emph{Bottom}: $A_V$ extinction map generated with 2MASS photometry
  with clusters overlaid.\label{fig-reg2} In both plots, B stars are
  marked with blue asterisks.}
\end{figure}

AFGL~2636 lies on the outskirts of Cygnus X North and exhibits similar
structure to the Diamond Ring. The clusters are displayed on an IRAC
channel 4 image in Figure~\ref{fig-reg2}. The largest and densest
cluster is G82.55+0.40 (red), which contains two B type stars. As with
the Diamond Ring region, we found 3 objects coincident with the red
cluster, whose optical photometry is consistent with a spectral type
earlier than B0 at the distance and reddening of Cygnus X. This
cluster has a Class II/I ratio of 12.5$\pm$3.9 which suggests that it
represents the oldest phase of star formation in this region. The cyan
cluster G82.40+0.23 which sits just below the red also has a high II/I
ratio of 12.0$\pm$12.5, but this large range makes its evolutionary
status uncertain. The other clusters have significantly lower II/I
ratios. The next largest cluster, G82.57+0.12 (orange) contains 18
Class I sources and has a II/I ratio of 1.6$\pm$0.5, making it likely
a much younger cluster. The nearby pink and purple clusters,
G82.61+0.06 and G82.58+0.21, are also relatively young, with II/I
ratios of 5.5$\pm$4.2 and 5.7$\pm$2.5, respectively. The gold cluster,
G82.40+0.23, contains 4 Class I sources and 7 Class II sources. It is
part of the long chain of Class I stars that forms a wide circle
around this region. The relative ages of the clusters suggest that
these latter objects are a later generation of star formation, which
is also consistent with their association with infrared dark clouds
(IRDCs) to the southeast of the AFGL~2636 cavity.

In the right panel of Figure~\ref{fig-reg2}, the clusters are shown
overlaid on the visual extinction map for this region. In contrast to
the Diamond Ring, all six clusters lie within areas of higher
extinction. However, we can see that the orange, purple, pink and gold
clusters reside in regions of higher extinction than the red and cyan
clusters. This trend is consistent with our conclusion that the red
and cyan clusters are the oldest in the region. The younger clusters
are still deeply embedded in molecular clouds, while the red and cyan
clusters have had time to remove some of their surrounding dust and
gas.

In AFGL~2636, there are 52 Class I stars and 214 Class II stars in
clusters, whereas, in the Diamond Ring, there are 65 Class I sources
and 618 Class II sources in clusters. Thus, the II/I ratio for all the
clusters in the Diamond Ring (9.5$\pm$1.2) is about twice that for
AFGL~2636 (4.1$\pm$0.64). We therefore conclude that AFGL~2636 is
likely to be younger than the Diamond Ring, which lies in the center
of Cygnus X North.

\subsection{Clustered Fractions}

We sought to find what fraction of young stars form in clusters as
opposed to in distributed populations. However, since defining cluster
membership is somewhat arbitrary, this question is difficult to
answer. Using the straight-line fit method with the cluster isolation
algorithm, we found that 67$\%$ of YSOs in the Diamond Ring region and
53$\%$ of YSOs in AFGL~2636 belong to clusters of at least 10
members. However, using a cutoff distance that maximizes the number of
clusters, we found that 58$\%$ and 63$\%$ of YSOs are in clusters in
the Diamond Ring and II, respectively. \citet{carpenter00} found that
$\sim$75\% of low mass stars form in clusters. Thus the regions in
Cygnus X, with clustered fractions of 53--67$\%$ fall slightly below
this level. This result may be more comparable to the work of Koenig
et al. (2008) who found that between 44 and 70$\%$ of YSOs in W5
reside in clusters with $\geq$10 members (depending on the choice of
clustering cutoff length). However, it should be noted that the study
of W5 covered a larger area than our smaller regions of Cygnus X, and
therefore included more distributed sources. The Diamond Ring region
and AFGL~2636 were chosen because they contain particularly dense
clusters around massive stars. If we looked at the entire Cygnus X
North region, the clustered fraction would likely be lower than
53--67$\%$.

\section{Conclusions and Future Work}

We have used optical spectroscopy from FAST and Hectospec to classify
possible OB sources in Cygnus X North. We classified a total of 536
sources and found a total of 24 B type stars and no O type stars,
although several candidate O stars are evident in the IPHAS survey's
photometric catalog. To verify that the spectrally observed stars are
indeed members of the Cygnus X complex, we plotted these sources on an
H-R diagram. We found that most of the stars seemed to fall on a line
consistent with a distance of 1.7~kpc, confirming their association
with Cygnus X.

We used \emph{Spitzer} IRAC and MIPS photometry to identify and
classify YSOs in Cygnus X North. We found 670 Class I, 7,249 Class II,
112 transition disk, and 200 embedded protostellar sources. We
identified two regions in Cygnus X North that contained a dense
grouping of YSOs around massive B stars. We used a minimal spanning
tree and cluster isolation technique to study the spatial distribution
of YSOs in these two regions. We found that 58--67$\%$ of YSOs belong
to clusters of $\geq$10 members in the Diamond Ring region and
53--63$\%$ of YSOs belong to clusters of $\geq$10 members in AFGL~2636.

We used the ratio of Class II/I sources in each cluster to approximate
a formation history of the clusters in each region assuming that a
higher ratio of Class II to Class I objects indicates a more evolved
(and thus older) cluster. We found that in both regions, a dense
cluster around one or more massive stars seems to have formed first. A
number of younger clusters are found surrounding the older
cluster. These younger clusters may have formed when stellar winds and
radiation from the most massive stars in the central cluster triggered
a new generation of star formation.

There remains much to be understood about the history of star
formation in the Cygnus X complex. By applying new algorithms for
defining cluster memberships, we can hope to better characterize the
structure of clusters in Cygnus X. As was mentioned earlier in this
paper, our FAST survey may have missed some of the brightest sources
in Cygnus X. Future studies will seek to locate and classify the most
massive stars in the region that were missed by our survey and will
allow us to better understand the evolution of star formation across
the entire Cygnus X region.

\acknowledgments

This work is based on observations made with the \emph{Spitzer Space
  Telescope}, which is operated by the Jet Propulsion Laboratory,
California Institute of Technology, under contract with NASA. This
work was supported in part by the National Science Foundation Research
Experiences for Undergraduates (REU) and Department of Defense Awards
to Stimulate and Support Undergraduate Research Experiences (ASSURE)
programs under Grant no. 0754568 and by the Smithsonian Institution.

Facilities: \facility{MMT(Hectospec)}, \facility{FLWO(FAST)}, \facility{2MASS ($JHK_S$)}, \facility{Spitzer (IRAC, MIPS)}

\clearpage
\appendix

\section{Spectral Sample}
Table~\ref{tbl-a1} lists the spectral classifications of the sources
for which we obtained FAST spectra. The source names were derived from
the position using the minutes and seconds of right ascension and
degrees and minutes of declination. The sources are listed with their
spectral types and errors, which were determined by SPTclass.

\begin{deluxetable}{lccccccc}
\tablecaption{Optical Spectroscopy Classifications}
\tablewidth{0pt}
\tablehead{\colhead{Source} & \colhead{$r'$} & \colhead{$r'$ error} & \colhead{$i'$} & \colhead{$i'$ error} & \colhead{Sp. Type} & \colhead{Type Err.} & \colhead{$A_V$}\\
\colhead{} & \colhead{(mag)} & \colhead{(mag)} & \colhead{(mag)} & \colhead{(mag)} & \colhead{} & \colhead{} & \colhead{(mag)}}

\startdata
J203534.33+414946.7 & 12.368 & 0.001 & 12.001 & 0.001 & G1.0 & 2.0 & 1.2 \\
J203535.86+414925.7 & 12.447 & 0.001 & 11.999 & 0.001 & G4.0 & 2.0 & 1.6 \\
J203538.92+415641.7 & 12.410 & 0.001 & 12.091 & 0.001 & F8.0 & 2.0 & 1.1 \\
J203548.37+420536.0 & 12.603 & 0.001 & 12.217 & 0.001 & F9.0 & 2.0 & 1.3 \\
J203556.64+414529.1 & 13.165 & 0.001 & 12.585 & 0.001 & F8.5 & 2.0 & 2.3 \\
J203614.10+422924.0 & 17.018 & 0.004 & 15.379 & 0.005 & K1.0 & 1.0 & 7.1 \\
J203615.46+420409.8 & 16.539 & 0.007 & 15.732 & 0.007 & K5.5 & 1.0 & 1.8 \\
J203626.20+414624.0 & 11.682 & 0.001 & 11.269 & 0.001 & F7.5 & 2.0 & 1.6 \\
J203626.62+423852.1 & 12.934 & 0.001 & 12.228 & 0.001 & B3.0 & 2.0 & 5.1 \\
J203629.11+420826.1 & 12.320 & 0.001 & 11.970 & 0.001 & F9.5 & 2.0 & 1.1 \\
\enddata
\tablecomments{This Table is published in its entirety in the
  electronic edition of the {\it Astrophysical Journal}.\label{tbl-a1}
  A portion is shown here for guidance regarding its form and
  content. The `Source' column gives the object name, which contains
  its Right ascension and Declination coordinates J2000.0.}
\end{deluxetable}


\begin{thebibliography}

\bibitem[{{Allen} {et~al.}(2007)}]{allen07} 
{Allen}, L. E. et al. 2007, Protostars and Planets V, eds. B. Reipurth, D. Jewitt, \& K. Keil (Univ. Arizona Press), 361

\bibitem[{{Battinelli} (1991)}]{batti91}
{Battinelli}, P. 1991, \aap, 244, 69

\bibitem[{{Carey} {et~al.}(2009)}]{carey09} 
{Carey}, S.~J., et al.\ 2009, \pasp, 121, 76

\bibitem[{{Carlson} {et~al.}(2007)}]{carlson07}
{Carlson}, L. R., et al.\ 2007, \apj, 665, 109

\bibitem[{{Carpenter} (2000)}]{carpenter00}
{Carpenter}, J.~M. 2000, \aj, 120, 3139

\bibitem[{{Cartwright} \& {Whitworth}(2004)}]{cartwright04}
{Cartwright}, A., \& {Whitworth}, A.~P. 2004, \mnras, 348, 589

\bibitem[{{Casertano} \& {Hut}(1985)}]{casertano85}
{Casertano}, S., \& {Hut}, P. 1985, \apj, 298, 80

\bibitem[{{Castor} {et~al.}(1975){Castor}, {McCray}, \& {Weaver}}]{castor75}
{Castor}, J., {McCray}, R., \& {Weaver}, R. 1975, \apj, 200, L107

\bibitem[{{Comer{\'o}n} {et~al.}(2008){Comer{\'o}n}, {Pasquali}, {Figueras}, \& {Torra}}]{comeron08}
{Comer{\'o}n}, F., {Pasquali}, A., {Figueras}, F., \& {Torra}, J. 2008, \aap, 486, 453

\bibitem[{{Currie} {et~al.}(2010)}]{currie10}
{Currie}, T., et al. 2010, arXiv:1002.1715

\bibitem[{{Downes} \& {Rinehart}(1966)}]{downes66}
{Downes}, D., \& {Rinehart}, R. 1966, \apj, 144, 937 

\bibitem[{{Drew} {et~al.}(2005){Drew}, {Greimel}, {Irwin}, {Aungwerojwit}, {Barlow}, {Corradi}, {Drake}, {G{\"a}nsicke}, {Groot}, {Hales}, {Hopewell}, {Irwin}, {Knigge}, {Leisy}, {Lennon}, {Mampaso}, {Masheder}, {Matsuura}, {Morales-Rueda}, {Morris}, {Parker}, {Phillipps}, {Rodriguez-Gil}, {Roelofs}, {Skillen}, {Sokoloski}, {Steeghs}, {Unruh}, {Viironen}, {Vink}, {Walton}, {Witham}, {Wright}, {Zijlstra}, \& {Zurita}}]{drew05}
{Drew}, J., et~al.\  2005, \mnras, 362, 753

\bibitem[{{Dutra} \& {Bica}(2001)}]{dutra01}
{Dutra}, C.~M., \& {Bica}, E. 2001, \aap, 376, 434

\bibitem[{{Elmegreen} \& {Lada}(1977)}]{elmegreen77}
{Elmegreen}, B.~G., \& {Lada}, C.~J. 1977, \apj, 214, 725

\bibitem[{{Fabricant} {et~al.}(1994){Fabricant}, {Hertz} \& {Szentgyorgyi}}]{fab94}
{Fabricant}, D., {Hertz}, E., \& {Szentgyorgyi}, A.~H. 1994, Proc. SPIE, 2198, 251

\bibitem[{{Fabricant} {et~al.}(1998)}]{fab98}
{Fabricant}, D. et al.\ 1998, \pasp, 110, 79

\bibitem[{{Fazio} {et~al.}(2004)}]{fazio04}
{Fazio}, G.~G. et al.\ 2004, \apj, 154, 10

\bibitem[{{Flaherty} {et~al.}(2007){Flaherty}, {Pipher}, {Megeath}, {Winston}, {Gutermuth}, {Muzerolle}, {Allen}, \& {Fazio}}]{flaherty07}
{Flaherty}, K.~M., {Pipher}, J.~L., {Megeath}, S.~T., {Winston}, E.~M., {Gutermuth}, R.~A., {Muzerolle}, J., {Allen}, L.~E., \& {Fazio}, G.~G. 2007, \apj, 663, 1069

\bibitem[{{Gehrz} {et~al.}(1980){Gehrz}, {Hackwell}, {Grasdalen}, {Merrill}, {Humphreys}, {Williamson}, {Puetter}, {Russell}, \& {Willner}}]{gehrz80} 
{Gehrz}, R.~D., et al.\ 1980, \apj, 85, 1071

\bibitem[{{Gonz{\'a}lez-Solares} {et~al.}(2008)}]{gonzalez08}
{Gonz{\'a}lez-Solares}, E.~A. et al.\ 2008, \mnras, 388, 89

\bibitem[{{Gutermuth} {et~al.}(2008)}] {gutermuth08} 
{Gutermuth}, R.~A. et al., 2008, \apj, 674, 336

\bibitem[{{Gutermuth} {et~al.}(2009)}] {gutermuth09}
{Gutermuth}, R.~A., {Megeath}, S.~T., {Myers}, P.~C., {Allen}, L.~E., {Pipher}, J.~L., \& {Fazio}, G.~G. 2009, \apjs, 184, 18

\bibitem[{{Hern{\'a}ndez} et al.(2004)}] {her04}
{Hern{\'a}ndez}, J., {Calvet}, N., {Brice{\~n}o}, C., {Hartmann}, L., {Berlind}, P. 2004, \apj, 127, 1682

\bibitem[{{Hern{\'a}ndez} (2005)}]{her05}
{Hern{\'a}ndez}, J. 2005, Ph.D. dissertation, Universidad de Los Andes, Merida, Venezuela

\bibitem[{{Hern{\'a}ndez} {et~al.}(2008){Hern{\'a}ndez}, {Hartmann}, {Calvet}, {Jeffries}, {Gutermuth}, {Muzerolle}, \& {Stauffer}}]{hern08}
{Hern{\'a}ndez}, J., {Hartmann}, L., {Calvet}, N., {Jeffries}, R.~D., {Gutermuth}, R., {Muzerolle}, J., \& {Stauffer}, J. 2008, \apj, 686, 1195

\bibitem[{{Hora} {et~al.}(2010)}]{hora10}
{Hora}, J.~L., et~al.\ 2010, in prep

\bibitem[{{Jordi} {et~al.}(2006){Jordi}, {Grebel}, \& {Ammon}}]{jordi06}
{Jordi}, K., {Grebel}, E.~K., \& {Ammon}, K. 2006, \aap, 460, 339

\bibitem[{{Kenyon} \& {Hartmann}(1995)}]{kenyon95}
{Kenyon}, S.~J., \& {Hartmann}, L. 1995, \apjs, 101, 117

\bibitem[{{Koenig} {et~al.}(2008)}]{koenig08} 
{Koenig}, X.~P., et~al.\ 2008, \apj, 688, 1142

\bibitem[{{Kn{\"o}dlseder} (2000)}]{knodl00}
{Kn{\"o}dlseder}, J. 2000, \aap, 360, 539

\bibitem[{{Lada} (1991)}]{lada91}
{Lada}, C.~J. 1991, in The Physics of Star Formation and Early Stellar Evolution, ed. C.~J. Lada \& N.~D. Kylafis (Dordrecht: Kluwer), 329

\bibitem[{{Lada} \& {Lada}(2003)}]{lada03} 
{Lada}, C.~J., \& {Lada}, E.~A. 2003, \araa, 41, 57

\bibitem[{{Lawrence} {et~al.}(2007){Lawrence}, {Warren}, {Almaini}, {Edge}, {Hambly}, {Jameson}, {Lucas}, {Casali}, {Adamson}, {Dye}, {Emerson}, {Foucaud}, {Hewett}, {Hirst}, {Hodgkin}, {Irwin}, {Lodieu}, {McMahon}, {Simpson}, {Smail}, {Mortlock}, \& {Folger}}]{law07}
{Lawrence}, A., et~al.\ 2007, \mnras, 379, 1599

\bibitem[{{Le Duigou} \& {Kn\"{o}dlseder}(2002)}]{leduigou02} 
{Le Duigou}, J.-M., \& {Kn\"{o}dlseder}, J. 2002, \aap, 392, 869

\bibitem[{{Lockman}(1989)}]{lockman89}
{Lockman}, F.~J. 1989, \apjs, 71, 469

\bibitem[{{Lucas} {et~al.}(2008){Lucas}, {Hoare}, {Longmore}, {Schr{\"o}der}, {Davis}, {Adamson}, {Bandyopadhyay}, {de Grijs}, {Smith}, {Gosling}, {Mitchison}, {G{\'a}sp{\'a}r}, {Coe}, {Tamura}, {Parker}, {Irwin}, {Hambly}, {Bryant}, {Collins}, {Cross}, {Evans}, {Gonzalez-Solares}, {Hodgkin}, {Lewis}, {Read}, {Riello}, {Sutorius}, {Lawrence}, {Drew}, {Dye}, \& {Thompson}}]{lucas08}
{Lucas}, P.~W., et~al.\ 2008, \mnras, 391, 136

\bibitem[{{Marston}(2004)}]{marston04}
{Marston}, A.~P. 2004, \apjs, 154, 333

\bibitem[{{Mizuno} {et~al.}(2008)}]{mizuno08} 
{Mizuno}, D.~R., et~al.\ 2008, \pasp, 120, 1028 

\bibitem[{{Piddington} \& {Minnett}(1952)}] {piddington52}
{Piddington}, J.~H., {Minnett}, H.~C., 1952, Aust. J. Sci. Res. A Phys. Sci., 5, 17

\bibitem[{{Povich} {et~al.}(2009){Povich}, {Churchwell}, {Bieging}, {Kang}, {Whitney}, {Brogan}, {Kulesa}, {Cohen}, {Babler}, {Indebetouw}, {Meade}, \& {Robitaille}, T.~P.}]{povich09}
{Povich}, M.~S., et~al.\ 2009, \apj, 696, 1278

\bibitem[{{Povich} \& {Whitney}(2010)}]{povich10}
{Povich}, M.~S., \& {Whitney}, B.~A. 2010, \apj, 714, L285

\bibitem[{{Reipurth} \& {Schneider}(2008)}]{reip08}
{Reipurth}, B., \& {Schneider}, N. 2008, in ASP, Handbook of Star Forming Regions, Vol. 1, ed. B. Reipurth (San Francisco, CA: ASP), 36

\bibitem[{{Rieke} {et~al.}(2004)}]{rieke04} 
{Rieke}, G.~H. et al.\ 2004, \apj, 154, 25

\bibitem[{{Robitaille} {et~al.}(2008){Robitaille}, {Meade}, {Babler}, {Whitney}, {Johnston}, {Indebetouw}, {Cohen}, {Povich}, {Sewilo}, {Benjamin}, \& {Churchwell}}]{robitaille08}
{Robitaille}, T.~P. et~al.\ 2008, \aj, 136, 2413

\bibitem[{{Ruch} {et~al.}(2007)}]{ruch07} 
{Ruch}, G.~T., {Jones}, T.~J., {Woodward}, C.~E., {Polomski}, E.~F., {Gehrz}, R.~D., \& {Megeath}, S.~T.\ 2007, \apj, 654, 338

\bibitem[{{Schlegel} {et~al.}(1998){Schlegel}, {Finkbeiner}, \& {Davis}}]{schlegel98}
{Schlegel}, D.~J., {Finkbeiner}, D.~P., \& {Davis}, M. 1998, \apj, 500, 525

\bibitem[{{Schneider} {et~al.}(2006)}]{schneider06}
{Schneider}, N., {Bontemps}, S., {Simon}, R., {Jakob}, H., {Motte}, F., {Miller}, M., {Kramer}, C., \& {Stutzki}, J. 2006, \aap, 458, 855

\bibitem[{{Schneider} {et~al.}(2007)}]{schneider07}
{Schneider}, N., {Simon}, R., {Bontemps}, S., {Comer{\'o}n}, F., \& {Motte}, F. 2007, \aap, 474, 873

\bibitem[{{Siess} {et~al.}(2000){Siess}, {Dufour}, \& {Forestini}}]{siess00}
{Siess}, L., {Dufour}, E., \& {Forestini}, M. 2000, \aap, 358, 593

\bibitem[{{Skrutskie} {et~al.}(2006){Skrutskie}, {Cutri}, {Stiening}, {Weinberg}, {Schneider}, {Carpenter}, {Beichman}, {Capps}, {Chester}, {Elias}, {Huchra}, {Liebert}, {Lonsdale}, {Monet}, {Price}, {Seitzer}, {Jarrett}, {Kirkpatrick}, {Gizis}, {Howard}, {Evans}, {Fowler}, {Fullmer}, {Hurt}, {Light}, {Kopan}, {Marsh}, {McCallon}, {Tam}, {Van Dyk}, \& {Wheelock}}]{skrut06}
{Skrutskie}, M.~F., et~al.\ 2006, \aj, 131, 1163

\bibitem[{{Str{\"o}mgren}(1939)}]{stromg39}
{Str{\"o}mgren}, B. 1939, \apj, 89, 526

\bibitem[{{Tokarz} \& {Roll}(1997)}]{tokarz97}
{Tokarz}, S.~P., \& {Roll}, J. 1997, in ASP Conf. Ser. 125, Astronomical Data and Software Systems VI, ed. G. Hunt \& H.~E. Payne (San Francisco:ASP), 140

\bibitem[{{Wendker}(1984)}]{wend84}
{Wendker}, H.~J. 1984, \aaps, 58, 291

\bibitem[{{Wendker} {et~al.}(1991){Wendker}, {Higgs}, \& {Landecker}}]{wend91}
{Wendker}, H.~J., {Higgs}, L.~A., \& {Landecker}, T.~L. 1991, \aap, 241, 551

\bibitem[{{Werner} {et~al.}(2004)}]{werner04} 
{Werner}, M.~W. et al. 2004, \apj, 154, 1

\bibitem[{{Zinnecker} \& {Yorke}(2007)}]{zinn07}
{Zinnecker}, H., \& {Yorke}, H.~W. 2007, \araa, 45, 481

\end{thebibliography}
\end{document}